\documentclass[reviewcopy]{elsarticle}

\usepackage[reviewcopy]{adndt}
\usepackage{longtable}
\usepackage{multirow}
\usepackage[pdfstartview=FitH,
            CJKbookmarks=true,
            bookmarksnumbered=true,
            bookmarksopen=true,
            colorlinks,
            pdfborder=001,
            linkcolor=blue,
            anchorcolor=blue,
            citecolor=blue
            ]{hyperref}

\usepackage{amsmath}

\usepackage{amssymb}

\biboptions{square,sort&compress}
\bibpunct[]{[}{]}{,}{n}{}{;}
\begin{document}

\begin{frontmatter}

\journal{Atomic Data and Nuclear Data Tables}

\title{Nuclear chiral doublet bands data tables}


\author{B.W. Xiong}
\address{State Key Laboratory of Nuclear Physics and Technology,
   School of Physics, Peking University, Beijing 100871, China.}

\author{Y.Y. Wang\corref{cor1}}
\address{School of Physics and Nuclear Energy Engineering and
   International Research Center for Nuclei and Particles in the
   Cosmos, Beihang University, Beijing 100191, China.}

\cortext[cor1]{flyyuan@buaa.edu.cn(Y.Y. Wang)}
\date{12.04.2018} 

\begin{abstract}
  Since the prediction of nuclear chirality in 1997, tremendous progresses
  both theoretically and experimentally have been achieved. Experimentally,
  59 chiral doublet bands in 47 chiral nuclei (including 8 nuclei with
  multiple chiral doublets) have been reported in $A\sim80,~100,~130$,~and~
  190 mass regions. The spins, parities, energies, ratios of the magnetic
  dipole transition strengths to the electric quadrupole transition
  strengths, and related references for these nuclei are compiled and listed
  in Table 1. For these nuclei with the magnetic dipole transition strengths
  and the electric quadrupole transition strengths measured, the
  corresponding results are given in Table 2. A brief discussion is provided
  after the presentation of energy $E$, energy difference $\Delta E$, energy
  staggering parameter $S(I)$, rotational frequency $\omega$, kinematic
  moment of inertia $\mathcal{J}^{(1)}$, dynamic moment of inertia
  $\mathcal{J}^{(2)}$, and ratio of the magnetic dipole transition strength
  to the electric quadrupole transition strength $B(M1)/B(E2)$ versus spin
  $I$ in each mass region.

\end{abstract}


\end{frontmatter}

\newpage

\tableofcontents
\listofDtables
\listofDfigures
\vskip5pc


\section{Introduction}

Chirality commonly exists in nature, such as the macroscopic spirals of
snail shells, the microscopic handedness of certain molecules, and human
hands~\cite{Meng2016Relativistic}. In geometry, a figure is chiral if it
cannot be mapped onto its mirror image by rotations and translations
alone. In particle physics, chirality is a dynamic property
distinguishing between the parallel and anti-parallel orientations of the
intrinsic spin with respect to the momentum of the massless particle. In
chemistry, the study of chirality is a very active topic appearing in
inorganic, organic, physical, biochemistry, and supramolecular chemistry.

\begin{figure}[h]
  \centering
  \includegraphics[height=4.8cm]{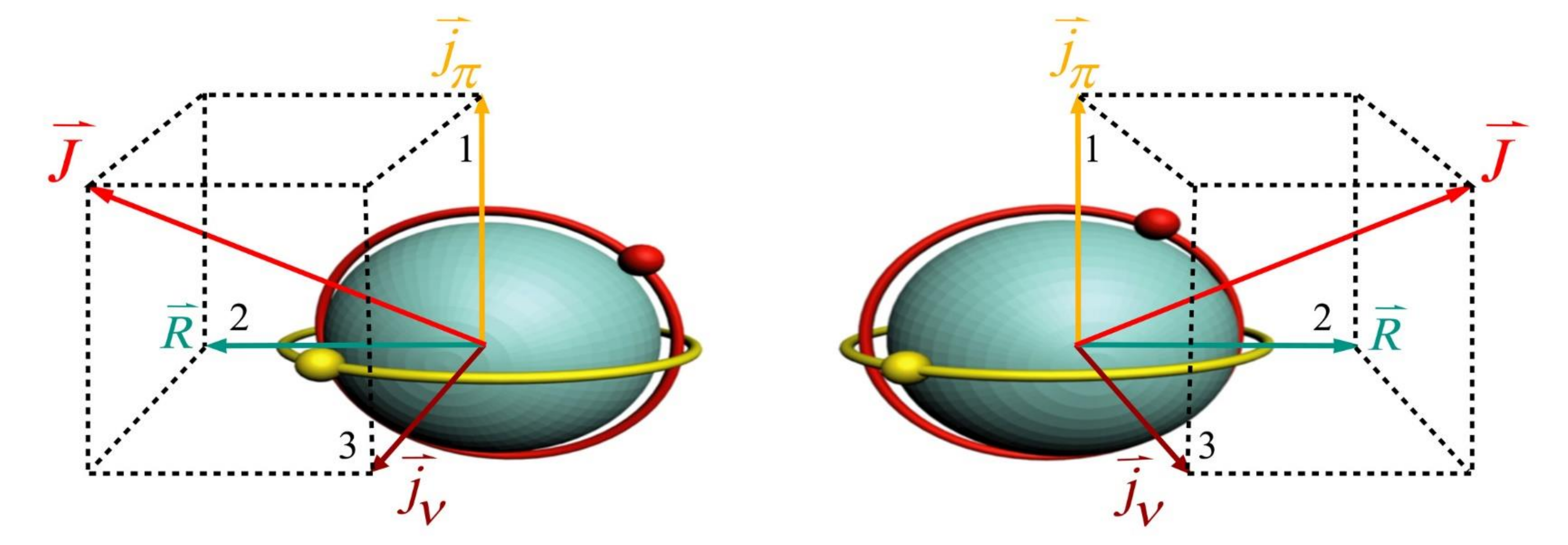}\\
  \caption{(Color online) Left- and right-handed chiral systems for a
     triaxial odd-odd nucleus \cite{Meng2010Open}.}
 \label{fig1}
\end{figure}

The chirality in nuclear physics was originally suggested by Frauendorf
and Meng in 1997~\cite{Frauendorf1997Tilted}. The physics mechanism of
nuclear chirality is illustrated in Fig. \ref{fig1}. For a rotational
nucleus with specific triaxial deformation, the collective angular
momentum favors alignment along the intermediate axis, which in this case
has the largest moment of inertia, while the angular momentum vectors of
the valence particles (holes) favor alignment along the nuclear short
(long) axis. The three mutually perpendicular angular momenta can be
arranged to form two systems with opposite chirality, namely left- and
right-handedness. These two systems are transformed into each other by
the chiral operator which combines time reversal and spatial rotation of
180$^\circ$,~$\chi = \mathcal{TR}(\pi)$. The spontaneous breaking of
chiral symmetry thus happens in the body-fixed reference frame. In the
laboratory reference frame, with the restoration of chiral symmetry due
to quantum tunneling, the so-called chiral doublet bands, i.e., a pair
of $\Delta I = 1$ bands (normally near degenerate) with the same parity,
are expected to be observed in triaxial nuclei
\cite{Frauendorf1997Tilted,Meng2010Open}.

The nuclear chirality, originally suggested in Ref.
\cite{Frauendorf1997Tilted} and vigorously investigated over the past
few years from both the theoretical and experimental standpoint,
continues to be the subject of intense discussion. The nuclear
chirality has become one of the hot topics in current nuclear physics
frontiers, as discussed in recent review articles~\cite{
Frauendorf2001Spontaneous,Meng2010Open,Meng2011Chirality,Meng2016Nuclear,
Zhou2016Multidimensionally,Sheikh2016Microscopic,RADUTA2016Specific,
Starosta2017Nuclear}.

Theoretically, the nuclear chirality is firstly predicted by particle
rotor model (PRM) and tilted axis cranking (TAC) approach in a single-$j$
shell \cite{Frauendorf1997Tilted}. Later on, various approaches have been
developed to describe the nuclear chiral doublet bands.

Depending on the number of valence nucleons and the orbits occupied,
1-particle-1-hole PRM \cite{Frauendorf1997Tilted,Peng2003Description,
Koike2004Chiral,Qi2009Examining,Qi2010Band, Chen2010Chiral},
2-quasiparticles PRM \cite{Zhang2007Chiral,Wang2008Description,
Wang2010Theoretical,Qi2011Chiral}, and $n$-particle-$n$-hole PRM
\cite{Qi2009Chirality,Qi2011Chirality,Qi2012Transition} have been developed.
The varieties of PRM include core quasiparticle coupling model~\cite{
Starosta2002Role,Koike2003Systematic,Droste2009Chiral}, interacting boson
fermion-fermion model \cite{Tonev2006,Tonev2007,Brant2008Dynamic,
Brant2009Chiral}, the generalized coherent state model \cite{Raduta2016New},
the angular momentum projection method \cite{Shimada2018Rotational}, and
pair truncated shell model \cite{Higashiyama2005New,Higashiyama2013Pair}.

The TAC adopted in Ref. \cite{Frauendorf1997Tilted} is based on a
single-$j$ mean field. Combining the spherical Woods-Saxon single-particle
energies and the deformed part of the Nilsson potential, chiral rotation
has been studied by the Strutinsky shell correction TAC method~\cite{
Dimitrov2000Chirality}. More microscopically, TAC based on covariant
density functional theory (CDFT) \cite{Madokoro2000Relativistic,
Meng2013Progress} has been introduced and applied to the studies of
chirality. The self-consistent Skyrme Hartree-Fock cranking model has also
been developed \cite{Olbratowski2004Critical}.

To go beyond mean field approximation to describe the chiral partners, one
can incorporate the quantum correlations by means of random phase
approximation \cite{Mukhopadhyay2007From,Almehed2011Chiral} or collective
Hamiltonian \cite{Chen2013Collective,Chen2016Two}. By taking into account
the quantum fluctuation along the collective degree of freedom, the
collective Hamiltonian goes beyond the mean field approximation and restores
the broken symmetry.

The attempts to understand the chiral doublet bands by the projected shell
model (PSM) \cite{Hara1995PROJECTED} have been performed in Ref. \cite{
Bhat2014Investigation}. Although the observed energy spectra and transitions
have been well reproduced in PSM, it is a big challenge to examine the
chiral geometry of angular momentum due to the complication that the
projected basis is defined in the laboratory frame and forms a nonorthogonal
set. Recently, the chiral geometry of the angular momentum is investigated
within the framework of PSM. The geometry of the angular momentum is
analyzed in terms of the distributions of its components on the three
intrinsic axes ($K~plot$) as well as the distributions of its tilted angles
in the intrinsic frame ($azimuthal~plot$) \cite{Chen2017Chiral,Chen2018APPB}.

The PRM is a quantal model consisting of the collective rotation and the
intrinsic single-particle motions, the energy splitting and quantum tunneling
between the doublet bands can be obtained directly. The rigid rotor with
quadrupole deformation parameters $\beta$ and $\gamma$ is assumed
\cite{Meng2010Open}.

Starting from an effective nucleon-nucleon interaction with Lorentz invariance,
the CDFT naturally includes the spin-orbit coupling and has achieved great
successes in describing many nuclear phenomena in stable and exotic nuclei of
the whole nuclear chart~\cite{Meng2016Relativistic,RING1996PPNP,MENG2006PPNP,
Meng2015JPG,LIANG2015Hidden}. It is interesting to search for nuclei with
triaxial deformation and configurations having not only one particle and one
hole but also several particles and several holes suitable for chirality in
CDFT in Ref. \cite{Meng2006Possible}, the adiabatic and configuration-fixed
constrained triaxial CDFT approaches are used to investigate the triaxial
shape coexistence and possible chiral doublet bands. A new phenomenon, the
existence of multiple chiral doublets (M$\chi$D), i.e., more than one pair of
chiral doublet bands in one single nucleus, is suggested for $^{106}$Rh. This
prediction remains with the time-odd fields included \cite{Yao2009Candidate},
and also holds true for other rhodium isotopes \cite{Peng2008Search}.

The first experimental evidence for M$\chi$D is reported in $^{133}$Ce in
2013 \cite{Ayangeakaa2013Evidence}. Later, a novel type of M$\chi$D with the
same configuration is reported in $^{103}$Rh \cite{Kuti2014Multiple}, which
shows that chiral geometry can be robust against the intrinsic excitation.
Then, two pairs of positive- and negative-parity doublet bands together with
eight strong electric dipole transitions linking their yrast positive- and
negative-parity bands, are identified in $^{78}$Br. This observation reports
the first example of chiral geometry in octupole soft nuclei and indicates
that nuclear chirality can be robust against the octupole correlations
\cite{Liu2016Evidence}. Recently, five pairs of nearly degenerate rotational
bands were identified in $^{136}$Nd \cite{Petrache2018Evidence}.

It should be mentioned that two pairs of chiral doublet bands in $^{105}$Rh
have been independently observed in Refs. \cite{
Alc2004Magnetic,Tim2004Experimental}, which have been confirmed to be
M$\chi$D by adiabatic and configuration-fixed constrained CDFT calculations
\cite{Li2011Multiple}. Similarly, one pair of chiral doublet bands in
$^{107}$Ag observed in Ref. \cite{He2012Quest}, which together with nearly
degenerate partner bands (though the difference in spin alignment in Fig.
\ref{fig16} needs further clarification) \cite{Dan1994Collective}, has been
claimed to show evidence of M$\chi$D \cite{Qi2013Possible}. Similarly, for
$^{138}$Nd, one pair of chiral doublet bands has been suggested in Ref.
\cite{Petrache2012Tilted}, which together with another pair of partner
bands, could be a new candidate of M$\chi$D \cite{Raduta2016New}.

Generally speaking, for the description of chiral rotations, three
dimensional tilted axis cranking CDFT (3D TAC-CDFT) is needed. The 3D
TAC-CDFT is firstly developed in Ref. \cite{Madokoro2000Relativistic}.
However, because of its numerical complexity, so far, it has been applied
only for the magnetic rotation in $^{84}$Rb. Focusing on the magnetic
rotation bands, in 2008, a completely new computer code for the
self-consistent 2D TAC-CDFT has been established \cite{Peng2008PRC}. It is
based on the non-linear meson-exchange models and includes considerable
improvements allowing systematic investigations. Based on a point-coupling
interaction, the 2D TAC-CDFT is developed to investigated the magnetic
rotation bands \cite{ZHAO2011PLB,Yu2012PRC}, antimagnetic rotation bands
\cite{Zhao2011PRL,Zhao2012PRC}, transitions of nuclear spin orientation
\cite{Zhao2015Impact,Wang2017Yrast}, and linear alpha cluster bands \cite{
Zhao2015PRL}, and demonstrates high predictive power \cite{
Meng2013Progress,Meng2016Nuclear}. Recently, the first applications of the
3D TAC-CDFT for nuclear chirality is reported in $^{106}$Rh \cite{
Zhao2017Multiple} and $^{136}$Nd \cite{Petrache2018Evidence}, respectively.

Originally the observation of two almost degenerate $\Delta I = 1$
rotational bands is considered as the fingerprint of chiral doublet bands
\cite{Frauendorf1997Tilted}. With the improvement of experimental
techniques, the lifetime measurements for chiral doublet bands become
possible. According to the selection rule for the electromagnetic
transition in ideal case, there occurs an alternation of stronger and
weaker $M$1 transitions with spin over the degenerate spin range of chiral
doublet bands, which can manifest as $B(M1)/B(E2)$ staggering as a function
of spin $I$. These characteristics are suggested as electromagnetic
transition fingerprints of chiral geometry in Ref. \cite{Koike2004Chiral}.
It is pointed out that in ideal chiral doublet bands, the corresponding
properties such as the identical or similarity in energies, spin alignments,
and electromagnetic transition probabilities, the spin independence of the
energy staggering parameters $S(I)$, and the staggering of $B(M1)/B(E2)$ are
summarized as fingerprints of ideal chiral doublet bands \cite{
Wang2007Examining}.

\begin{figure}[h]
  \centering
  \includegraphics[height=8.0cm]{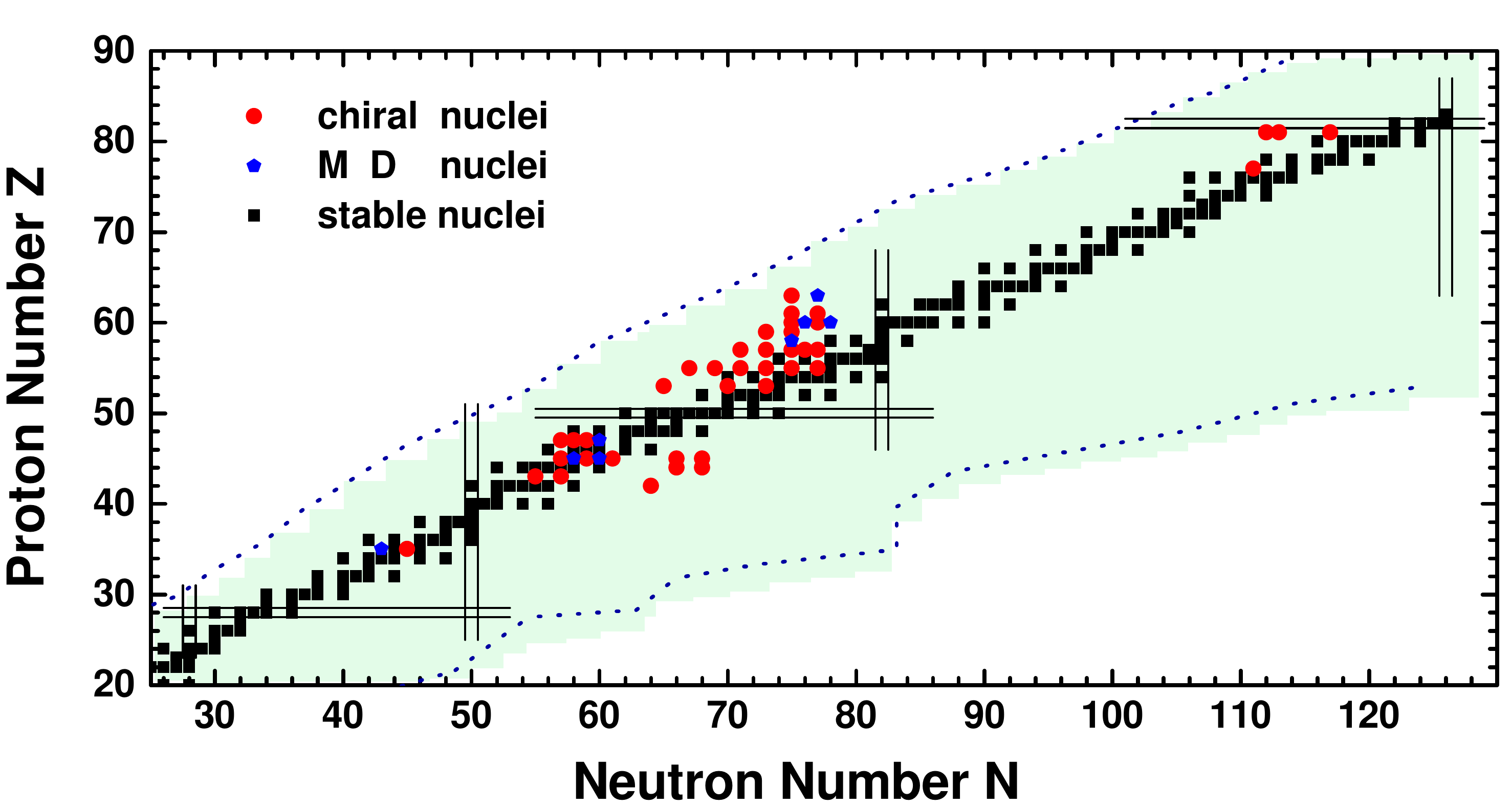}\\
  \caption{(Color online) The nuclides with chiral doublet bands (red
    circles) and M$\chi$D (blue pentagons) observed in the nuclear chart.
    The black squares represent stable nuclides.}
  \label{fig2}
\end{figure}

During the last two decades, lots of experimental efforts have been devoted
to search for nuclear chirality. Up to now, 59 chiral doublet bands in 47
chiral nuclei (including 8 M$\chi$D nuclei) have been reported in the
$A\sim$~80~\cite{Liu2016Evidence,Wang2011The}, 100~\cite{Zhu2009Search,
Ding2010Proposed,Joshi2005First,Y2009ODD,Tonev2014Candidates,
Kuti2014Multiple,Vaman2004Chiral,Alc2004Magnetic,Tim2004Experimental,
Joshi2004Stability,Luo2004Level,Wang2013High,Tim2007High,
Lieder2014Resolution,He2012Quest}, 130~\cite{Starosta2001Chirality,
Zhao2009Observation,Zheng2012Abnormal,Yonnam2005,Selvakumar2015Evidence,
Grodner2011Partner,Grodner2006128Cs,Simons2005Evidence,
Rainovski2003Candidate,Ma2012Candidate,Koike2001Observation,Kuti2013Medium,
Petrache2016Triaxial,Bark2001Candidate,Ayangeakaa2013Evidence,
Timar2011Medium,Mukhopadhyay2007From,Petrache2018Evidence,Petrache1997High,
Petrache2012Tilted,Hartley2001Detailed,Ma2018Candidate,Hecht2001Evidence,
Hecht2003Evidence}, and 190~\cite{Balabanski2004Possible,
Ndayishimye2017Chiral,Masiteng2014Rotational,Lawrie2010Candidate} mass
regions. The distribution of the observed chiral nuclei in the nuclear chart
is given in Fig. \ref{fig2}.

It should be pointed out that the lifetime measurements which are essential
to extract the absolute electromagnetic transition probabilities are still
rare for the chiral nuclei candidates.

In order to promote the study of chiral symmetry in atomic nuclei, the
compilations of the data for chiral doublet bands are highly demanded. This
is the purpose of the present paper.

In section \ref{Systematics of chiral doublet bands}, the figures are
presented for energy $E$, energy difference $\Delta E$, energy staggering
parameter $S(I)$, rotational frequency $\omega$,  kinematic moment of
inertia $\mathcal{J}^{(1)}$, dynamic moment of inertia $\mathcal{J}^{(2)}$,
and ratio of the magnetic dipole transition strength to the electric
quadrupole transition strength $B(M1)/B(E2)$ versus spin $I$ in each mass
region. The explanation of the tables for chiral doublet bands is followed.
Finally, a brief summary is given.

\clearpage

\section{Systematics of chiral doublet bands\label{Systematics of chiral
doublet bands}}

\subsection{Energy spectra \label{Energy spectra}}

\begin{figure}[htbp]
  \centering
  \includegraphics[height=5.8cm]{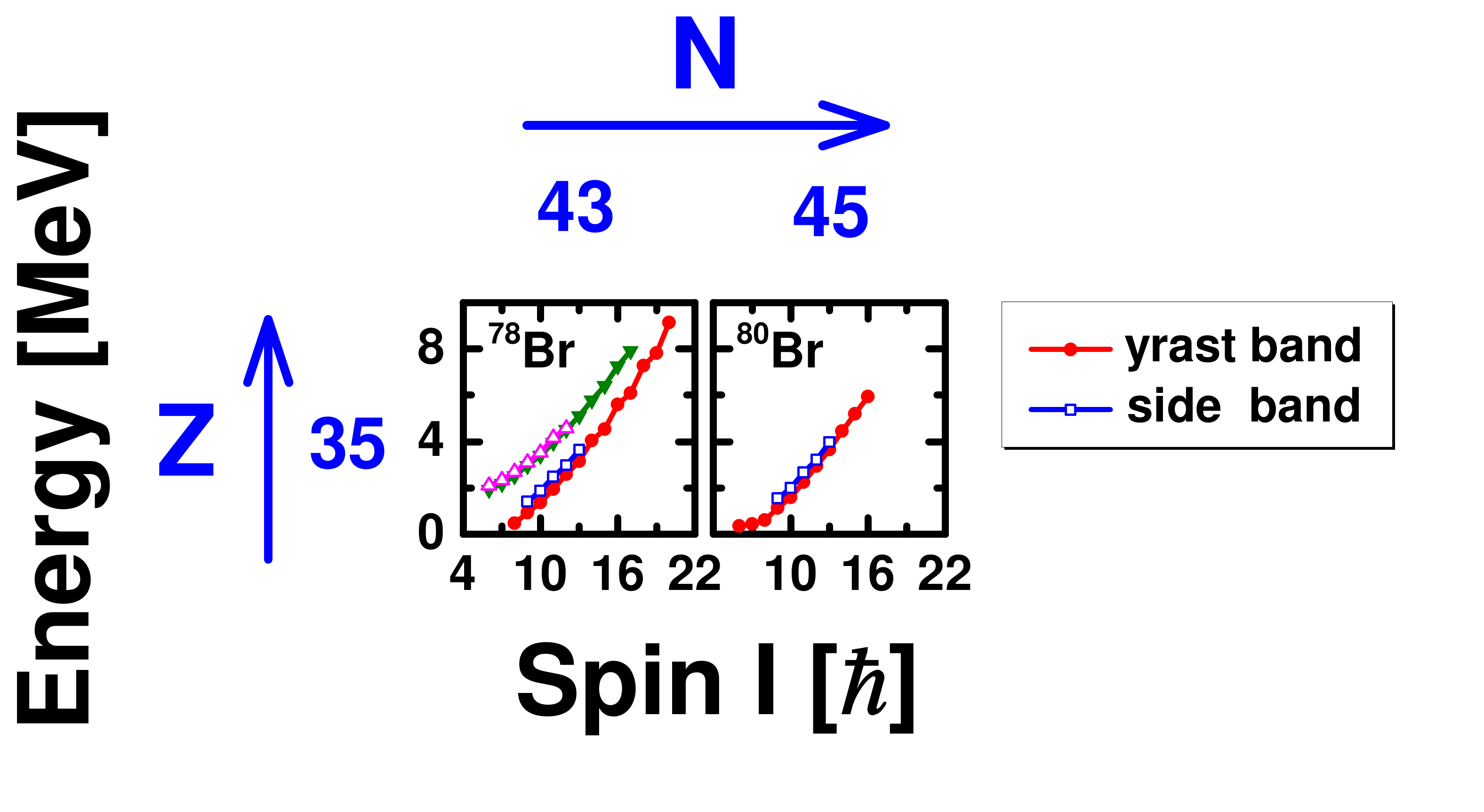}\\
  \caption{(Color online) Energies versus spin for chiral doublet bands in
    $A\sim$~80 mass region. The existence of M$\chi$D is suggested in
    $^{78}$Br. The excited chiral doublet bands are shifted by 1.5~MeV.}
  \label{fig3}
\end{figure}

\begin{figure}[htbp]
  \centering
  \includegraphics[height=11.5cm]{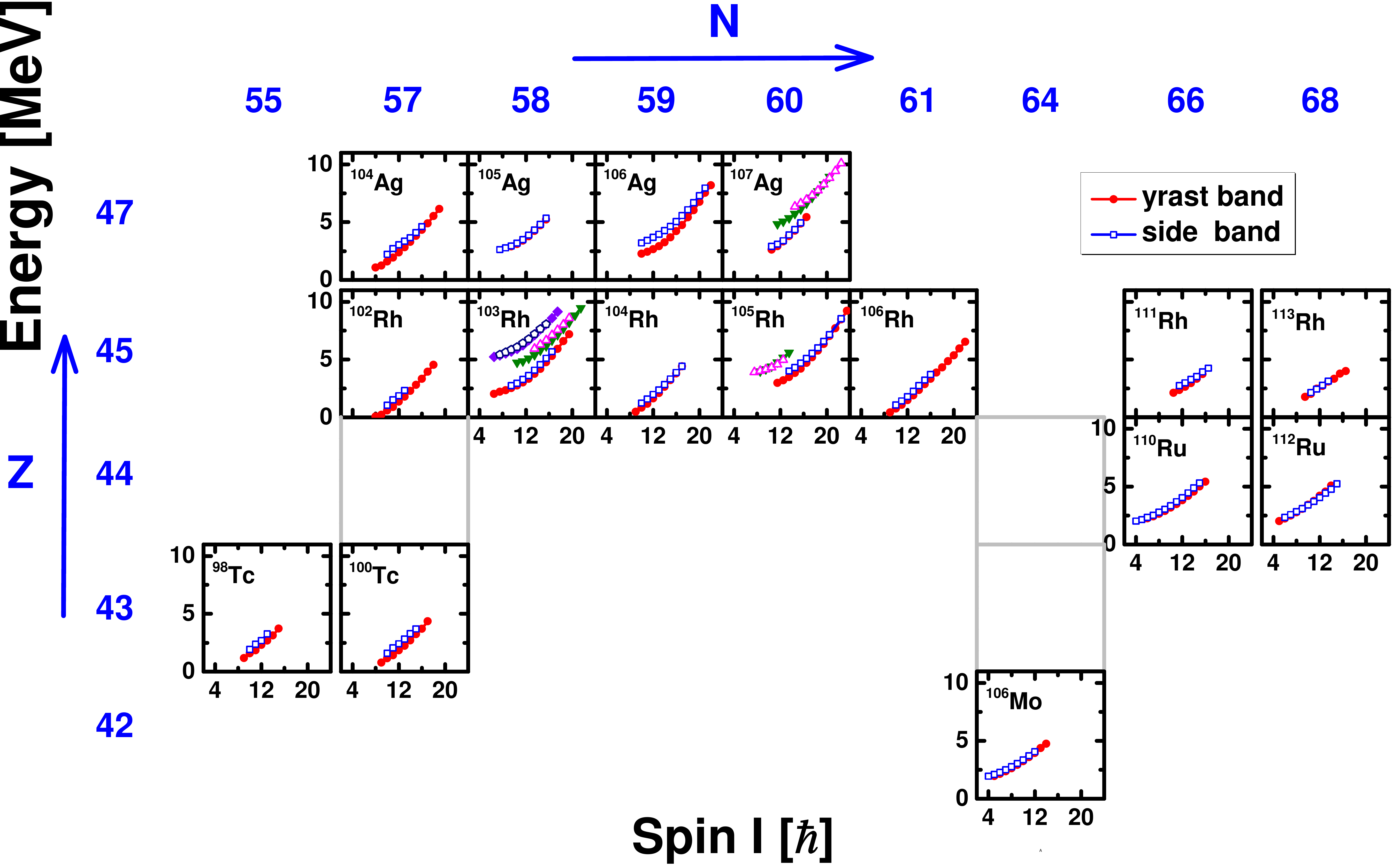}\\
  \caption{(Color online) Energies versus spin for chiral doublet bands in
    $A\sim$~100 mass region. The existence of M$\chi$D are suggested in
    $^{103}$Rh, $^{105}$Rh, and $^{107}$Ag. One pair of excited bands in $^{103}$Rh are shifted by 1.5 MeV, and another are shifted by 3 MeV.
    The excited chiral doublet bands in $^{105}$Rh and $^{107}$Ag are
    shifted by 1.5~MeV. The gray boxes are used to connect the various
    parts of this figure.}
  \label{fig4}
\end{figure}

\begin{figure}[htbp]
  \centering
  \includegraphics[height=15.5cm]{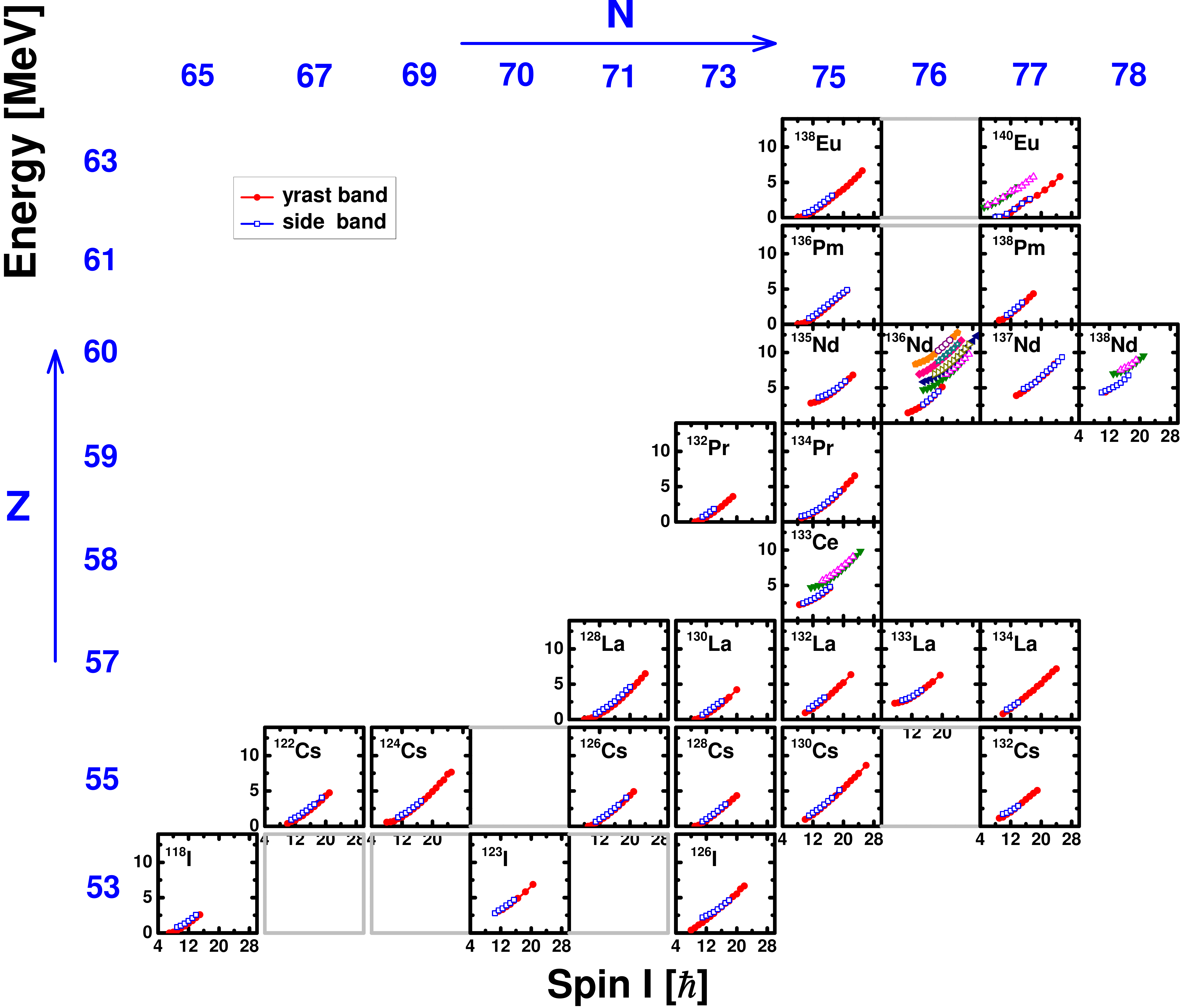}\\
  \caption{(Color online) Energies versus spin for chiral doublet bands in
    $A\sim$~130 mass region. The existences of M$\chi$D are suggested in
    $^{133}$Ce, $^{136}$Nd, $^{138}$Nd, and $^{140}$Eu. For $^{136}$Nd,
    there are five pairs of chiral doublet bands, the two lowest ones are
    shift by -3.0~MeV and -1.5~MeV, respectively, and the next two higher
    ones are shift by 1.5~MeV and 3.0~MeV, respectively. For other
    M$\chi$D, the excited chiral doublet bands are shifted by 1.5~MeV. The
    gray boxes are used to connect the various parts of this figure.}
  \label{fig5}
\end{figure}

\begin{figure}[htbp]
  \centering
  \includegraphics[height=9cm]{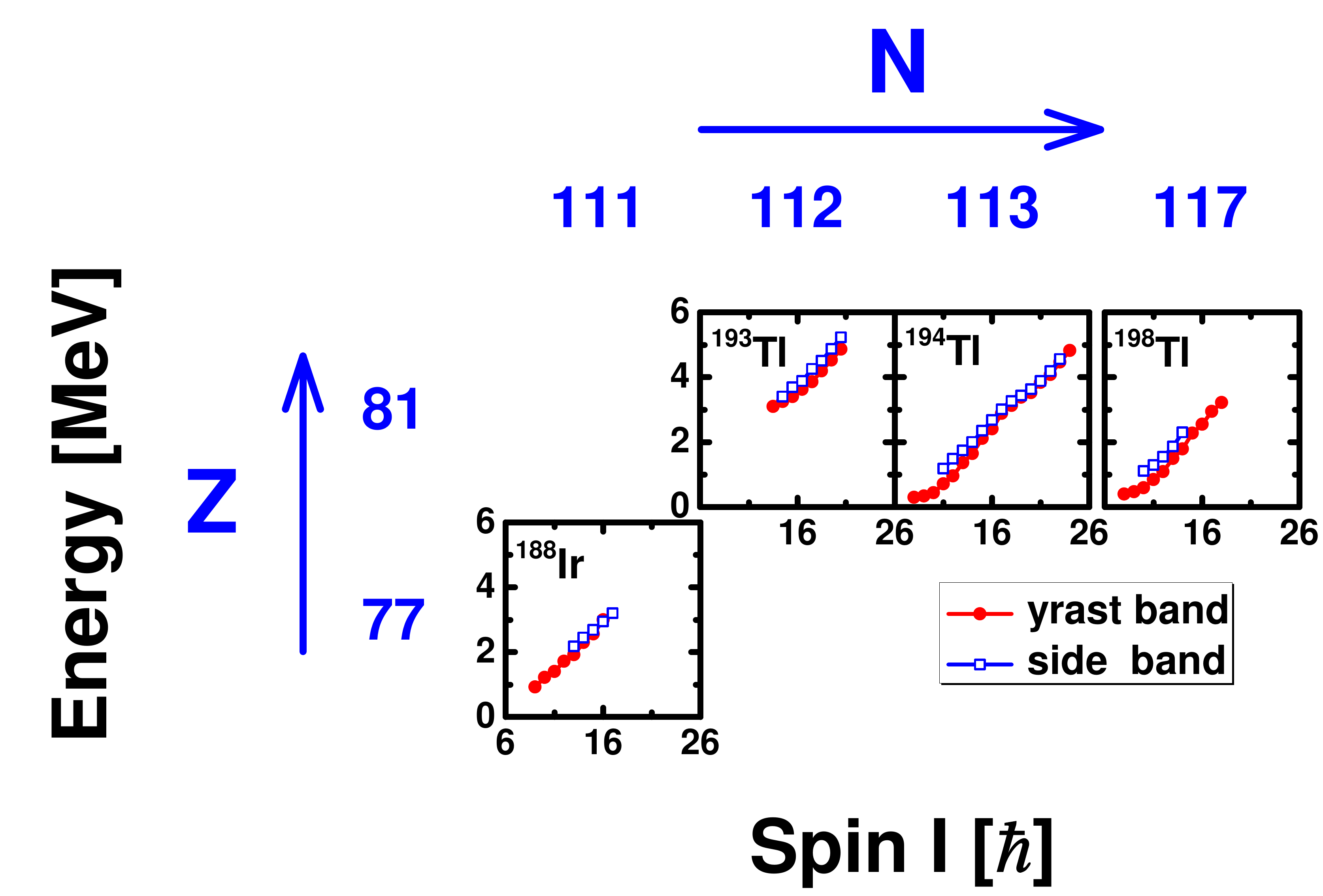}\\
  \caption{(Color online) Energies versus spin for chiral doublet bands in
    $A\sim$~190 mass region.}
  \label{fig6}
\end{figure}

The energy spectra for all chiral doublet bands in $A\sim80,~100,~130$,
and~190 mass regions are given in Fig. \ref{fig3}-\ref{fig6}, respectively.
For $\mathrm{M\chi D}$ in $^{78}$Br, $^{105}$Rh, $^{107}$Ag, $^{133}$Ce,
$^{138}$Nd, and $^{140}$Eu, the excited chiral doublet bands are shifted
by 1.5 MeV. For $\mathrm{M\chi D}$ in $^{103}$Rh, there are three pairs of
chiral doublet bands, one pair of excited bands are shifted by 1.5 MeV, and
another are shifted by 3 MeV. As for $\mathrm{M\chi D}$ in $^{136}$Nd,
there are five pairs of chiral doublet bands, the two lowest ones are shift
by -3.0~MeV and -1.5~MeV, respectively, and the next two higher ones are
shift by 1.5~MeV and 3.0~MeV, respectively.

The fingerprint for chiral doublet bands in energy, i.e., two almost
degenerate $\Delta I = 1$ rotational bands, is demonstrated in the figures.

The nuclear chirality occurs at the lowest spin 4$\hbar$ in $^{106}
\mathrm{Mo}$ and $^{110}\mathrm{Ru}$, and the highest spin 29$\hbar$ in
$^{136}\mathrm{Nd}$. The bands in $^{106}\mathrm{Mo}$ and $^{110}
\mathrm{Ru}$ are interpreted as soft chiral vibration~\cite{Zhu2005Soft,
Luo2009Evolution,Y2009ODD}. In $^{136}\mathrm{Nd}$, the chiral rotations
were interpreted by TAC-CDFT with the assigned configurations
\cite{Petrache2018Evidence}.

Generally, the energies for the partners are close to each other and tend
to be almost completely degenerate at high spins. There are several
exceptions: for nuclei $^{112}$Ru, $^{105}$Rh, $^{107}$Ag, $^{136}$Nd, $^{137}$Nd, $^{140}$Eu, and $^{188}$Ir, crossings between some partner 
bands exist.

\clearpage

\subsection{Energy difference}

\begin{figure}[htbp]
  \centering
  \includegraphics[height=5.1cm]{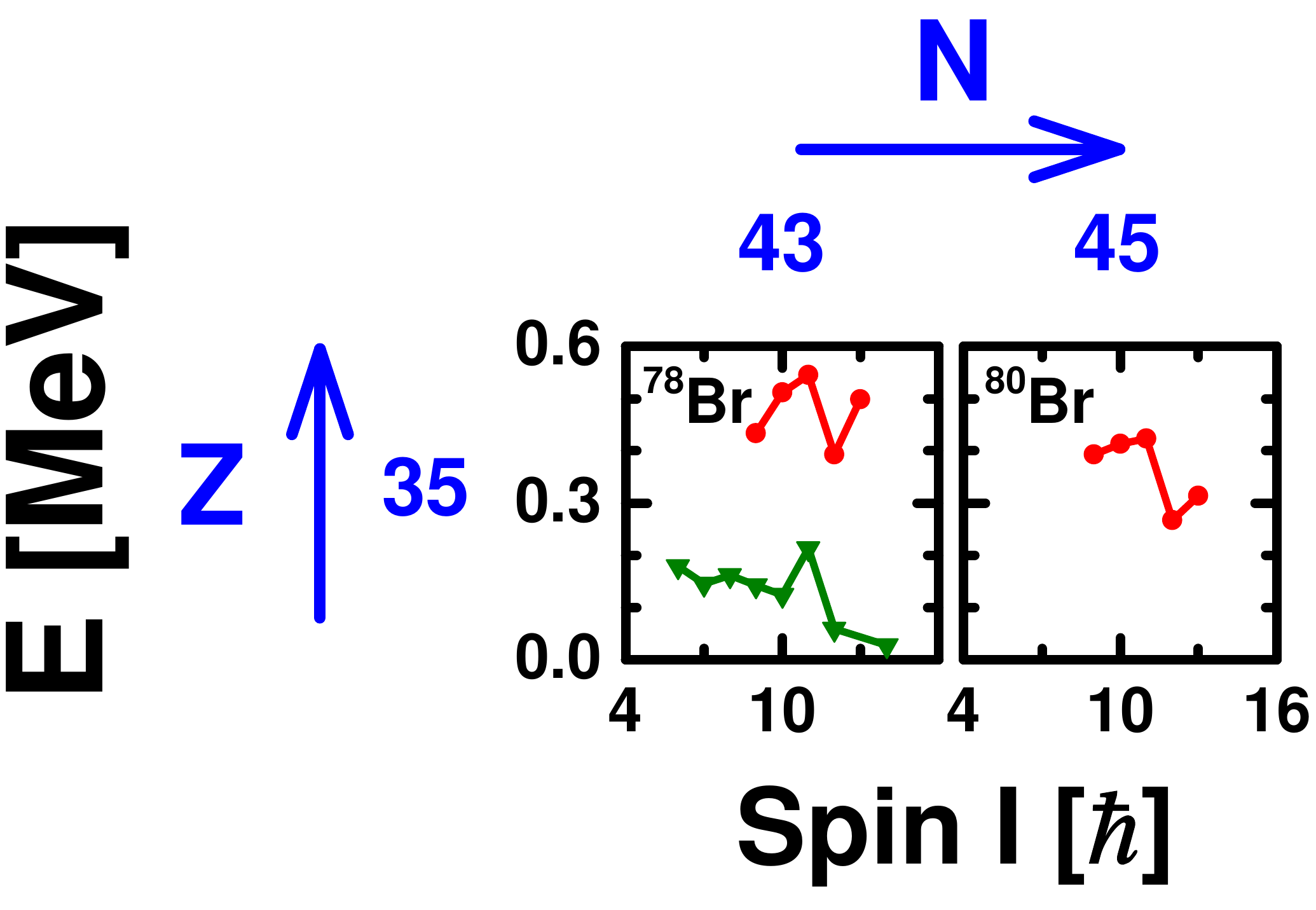}\\
  \caption{(Color online) Energy differences between the chiral doublet bands
    versus spin in $A\sim$~80 mass region. The red and green colors represent
    the energy differences between the chiral partners and the excited chiral
    partners, respectively.}
  \label{fig7}
\end{figure}

\begin{figure}[htbp]
  \centering
  \includegraphics[height=11.5cm]{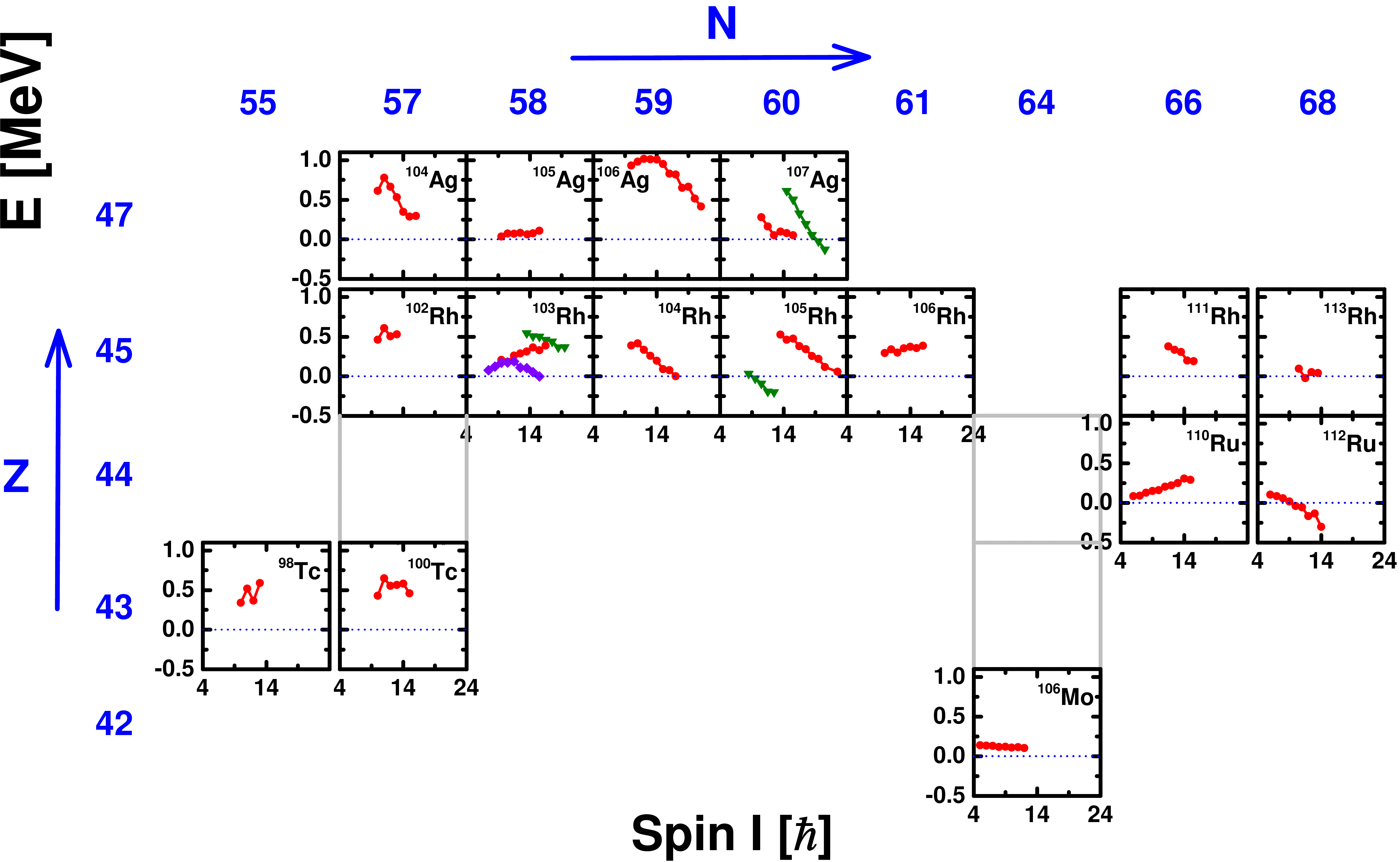}\\
  \caption{(Color online) Energy differences between the chiral doublet bands
    versus spin in $A\sim$~100 mass region.}
  \label{fig8}
\end{figure}

\begin{figure}[htbp]
  \centering
  \includegraphics[height=15.5cm]{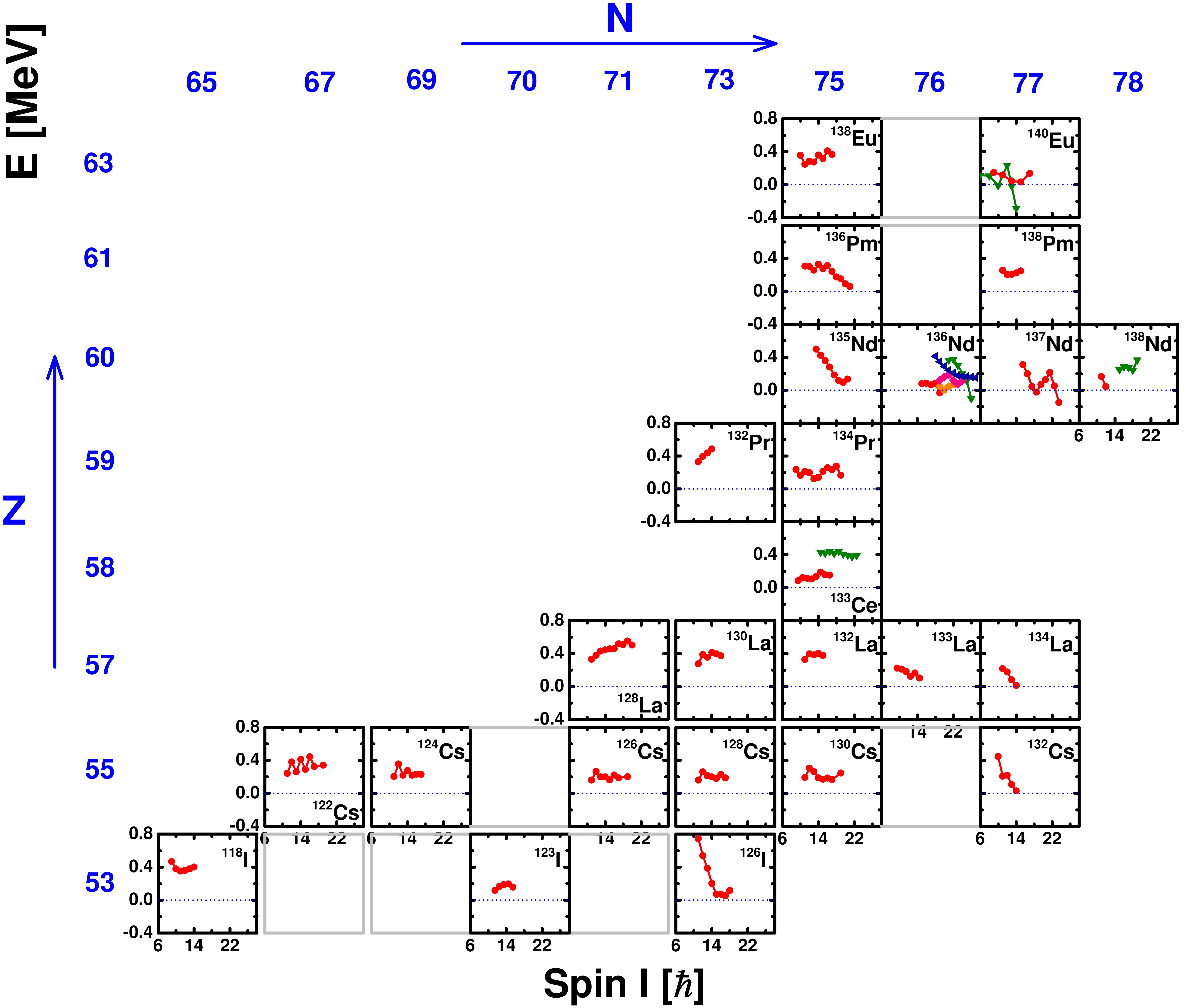}\\
  \caption{(Color online) Energy differences between the chiral doublet bands
    versus spin in $A\sim$~130 mass region.}
  \label{fig9}
\end{figure}

\begin{figure}[htbp]
  \centering
  \includegraphics[height=7.5cm]{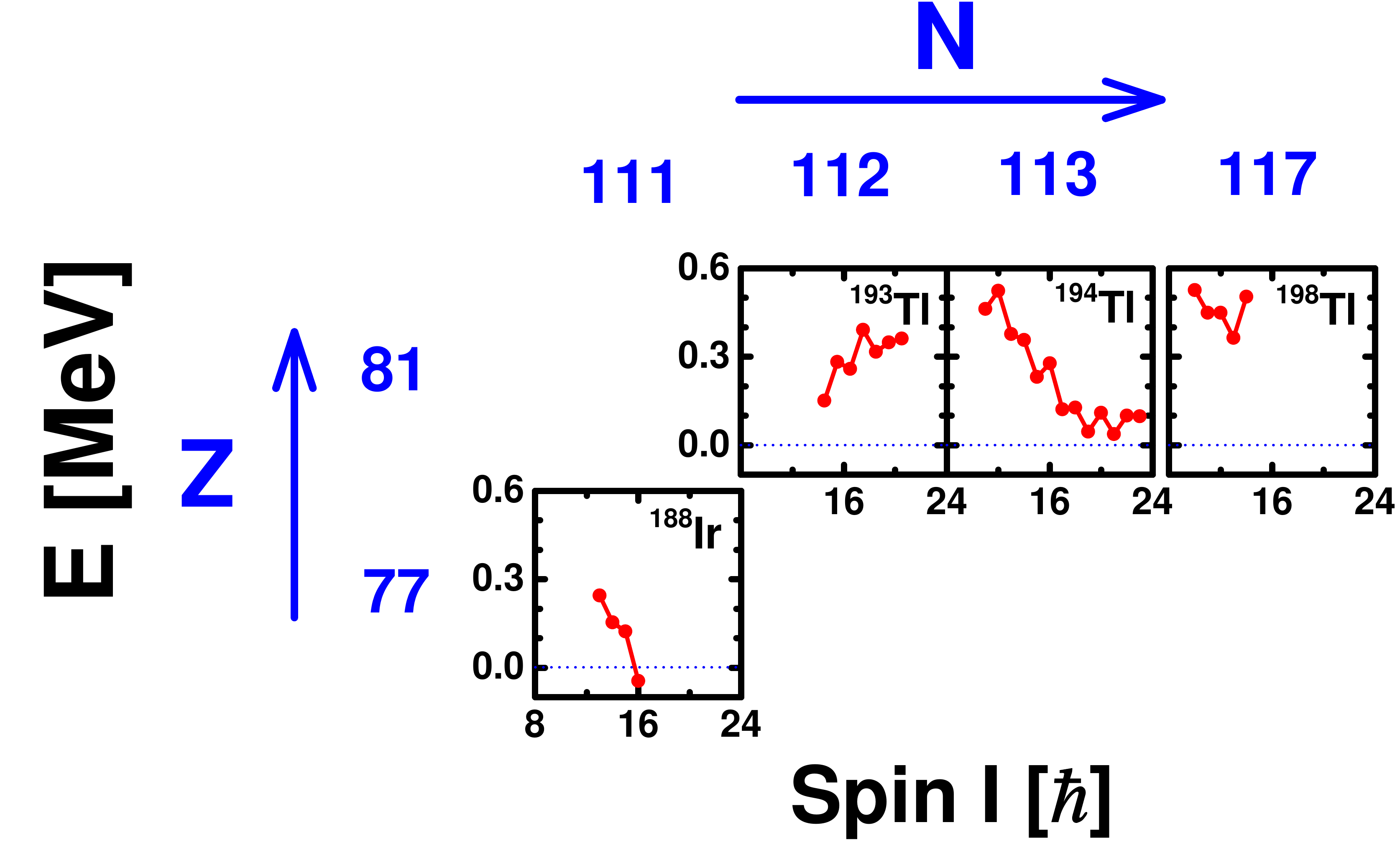}\\
  \caption{(Color online) Energy differences between the chiral doublet bands
    versus spin in $A\sim$~190 mass region.}
  \label{fig10}
\end{figure}

The energy differences $\Delta E(I)=E_{\mathrm{side}}(I)-E_{\mathrm{yrast}}
(I)$ between yrast band and side band for all chiral doublet bands in
$A\sim80,~100, ~130$, and~190 mass regions are given in Figs. \ref{fig7}-
\ref{fig10}, respectively. Although the chiral partner bands have energies
close to each other, it is rare to observe a crossing between them. If
crossing occurs, carefully examinations for chirality are necessary, as has
been done in $^{106}$Ag \cite{Lieder2014Resolution}.

The energy differences $\Delta E$ are below 600 keV for all chiral doublet
bands except for $^{100}$Tc, $^{102}$Rh, $^{104}$Ag, $^{106}$Ag, $^{107}$Ag,
and $^{126}$I. Several negative values exist for $\Delta E$ in nuclei
$^{112}$Ru, $^{105}$Rh, $^{107}$Ag, $^{136}$Nd, $^{137}$Nd, $^{140}$Eu, and
$^{188}$Ir which correspond to the crossing between the partner bands in
Section \ref{Energy spectra}.

\clearpage

\subsection{Energy staggering}

\begin{figure}[htbp]
  \centering
  \includegraphics[height=5.8cm]{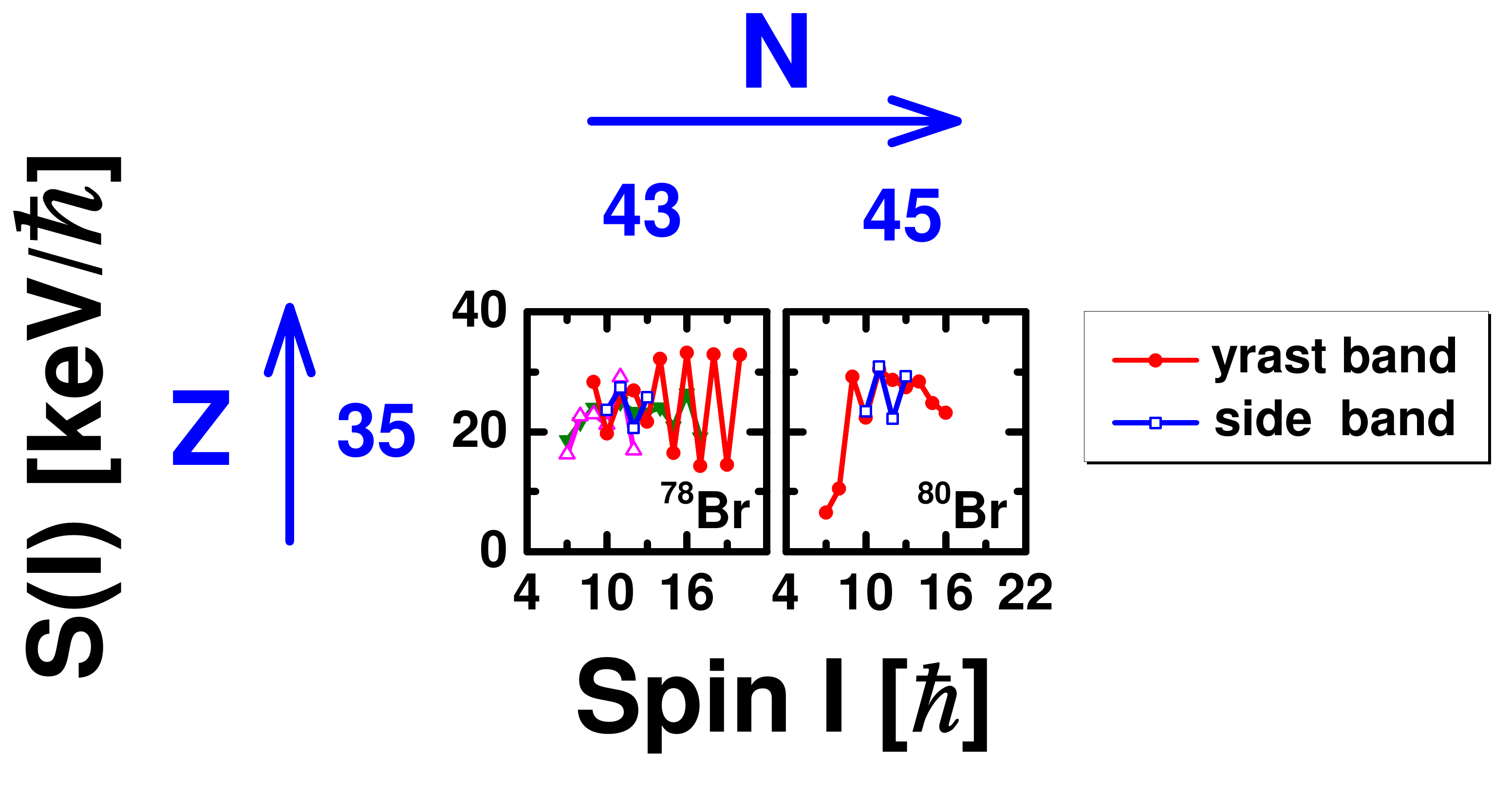}\\
  \caption{(Color online) Energy staggering parameters as functions of spin
    in $A\sim$~80 mass region.}
  \label{fig11}
\end{figure}

\begin{figure}[htbp]
  \centering
  \includegraphics[height=11.5cm]{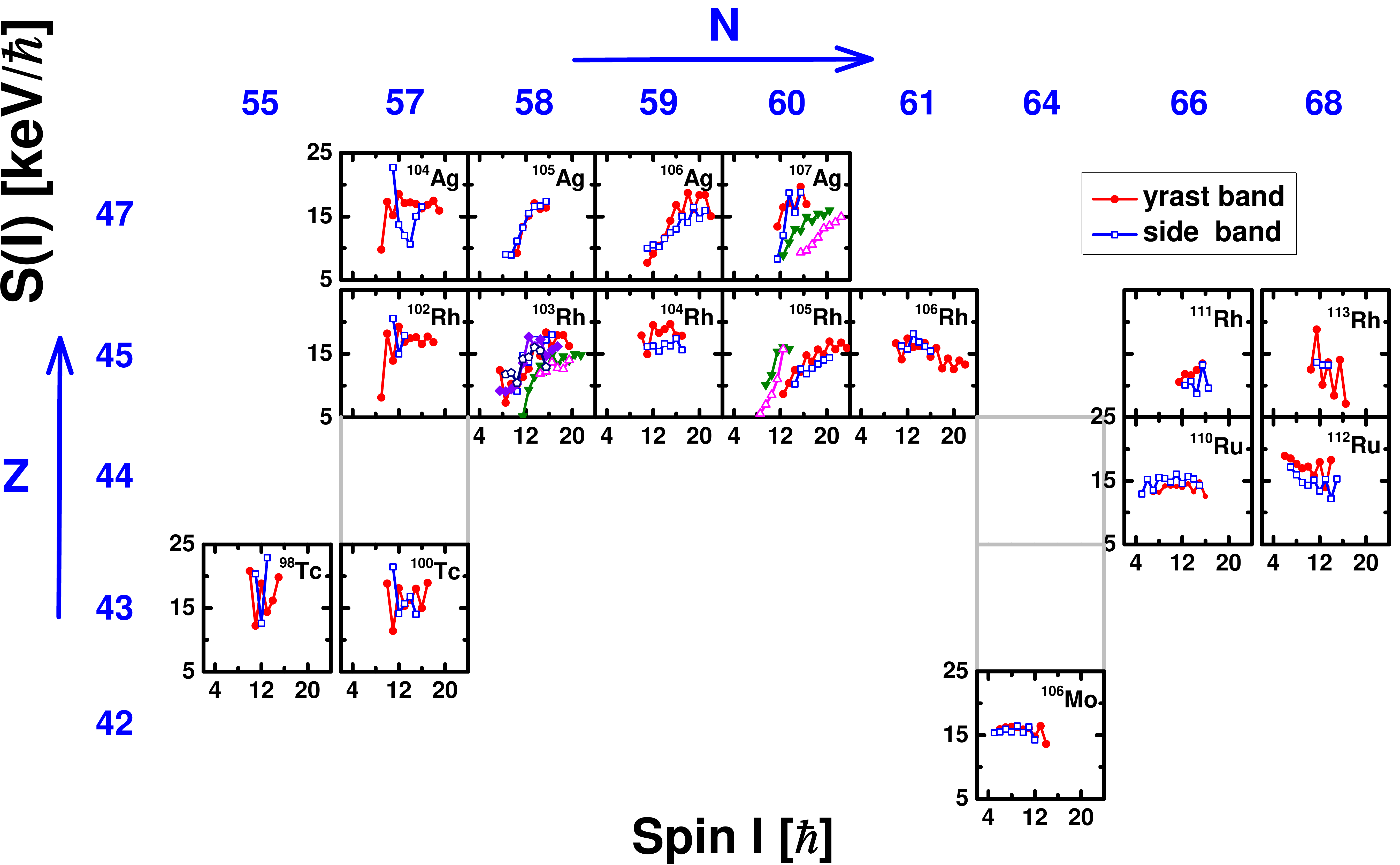}\\
  \caption{(Color online) Energy staggering parameters as functions of spin
    in $A\sim$~100 mass region.}
  \label{fig12}
\end{figure}

\begin{figure}[htbp]
  \centering
  \includegraphics[height=15.5cm]{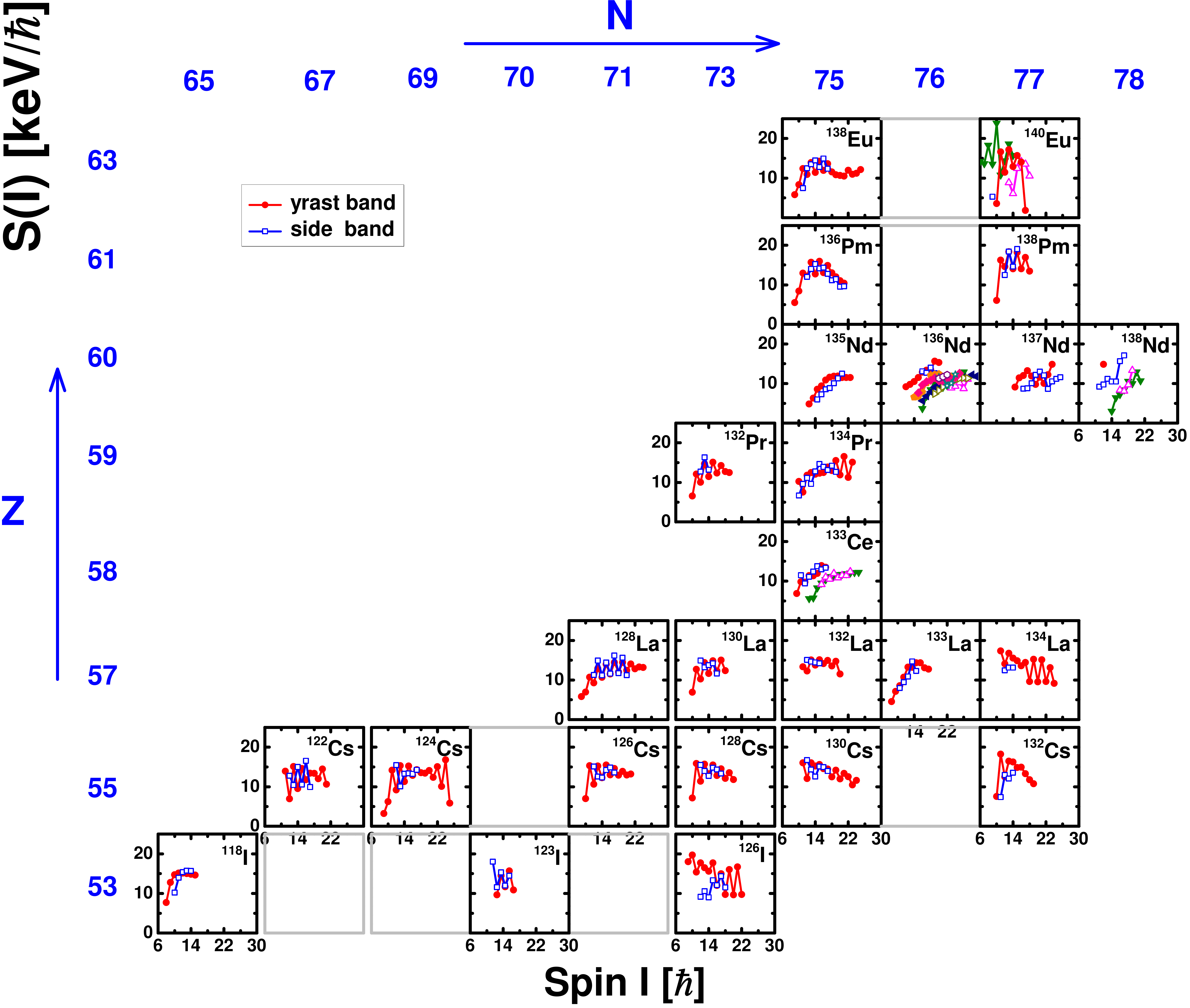}\\
  \caption{(Color online) Energy staggering parameters as functions of spin
    in $A\sim$~130 mass region.}
  \label{fig13}
\end{figure}

\begin{figure}[htbp]
  \centering
  \includegraphics[height=9cm]{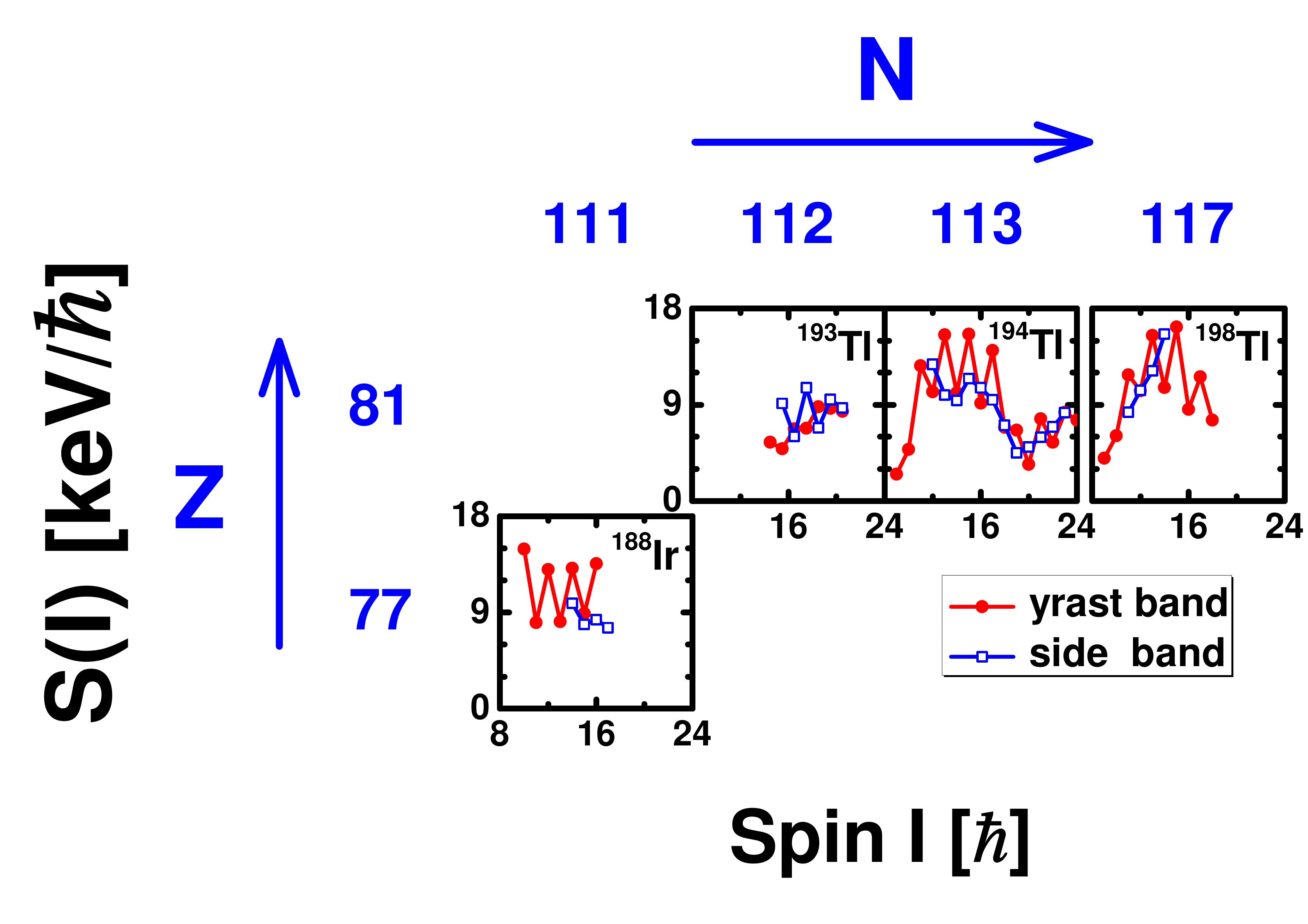}\\
  \caption{(Color online) Energy staggering parameters as functions of spin
    in $A\sim$~190 mass region.}
  \label{fig14}
\end{figure}

From the energy staggering parameter defined as $S(I)=[E(I)-E(I-1)]/2I$, the
energy staggering parameters as functions of spin for all chiral doublet bands
in $A\sim$~80, 100,~130,~and 190 mass regions are given in Figs.
\ref{fig11}-\ref{fig14}, respectively.

For the ideal chiral doublet bands, the $S(I)$ values should possess a smooth
dependence with spin, and it has been taken as a possible fingerprint.

Normally, the values of $S(I)$ change dramatically at the band head. At
certain spin range, they show a smooth dependence with spin. The change of
$S(I)$ values is around 20 keV/$\hbar$ in $A\sim80$ mass region, and decrease
to 10 keV/$\hbar$ for most nuclei in $A\sim100,~130$, and~190 mass regions,
except for $^{140}$Eu, $^{194}$Tl, and $^{198}$Tl whose change is around 20
keV/$\hbar$.

\clearpage

\subsection{Rotational frequency}

\begin{figure}[htbp]
  \centering
  \includegraphics[height=5.8cm]{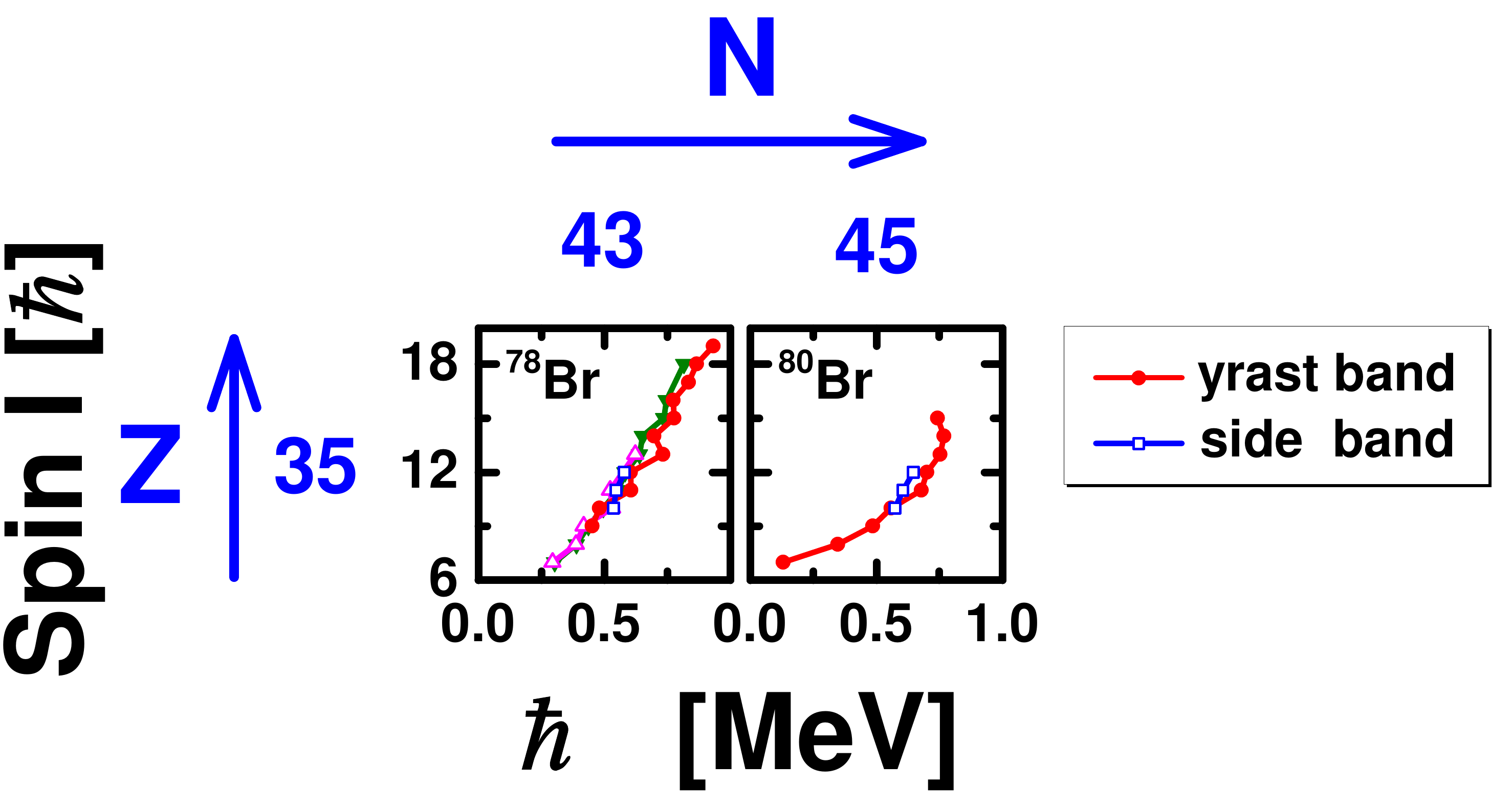}\\
  \caption{(Color online) The relation between the spin and the
   rotational frequency for chiral doublet bands in $A\sim$~80 mass
   region.}
  \label{fig15}
\end{figure}

\begin{figure}[htbp]
  \centering
  \includegraphics[height=11.5cm]{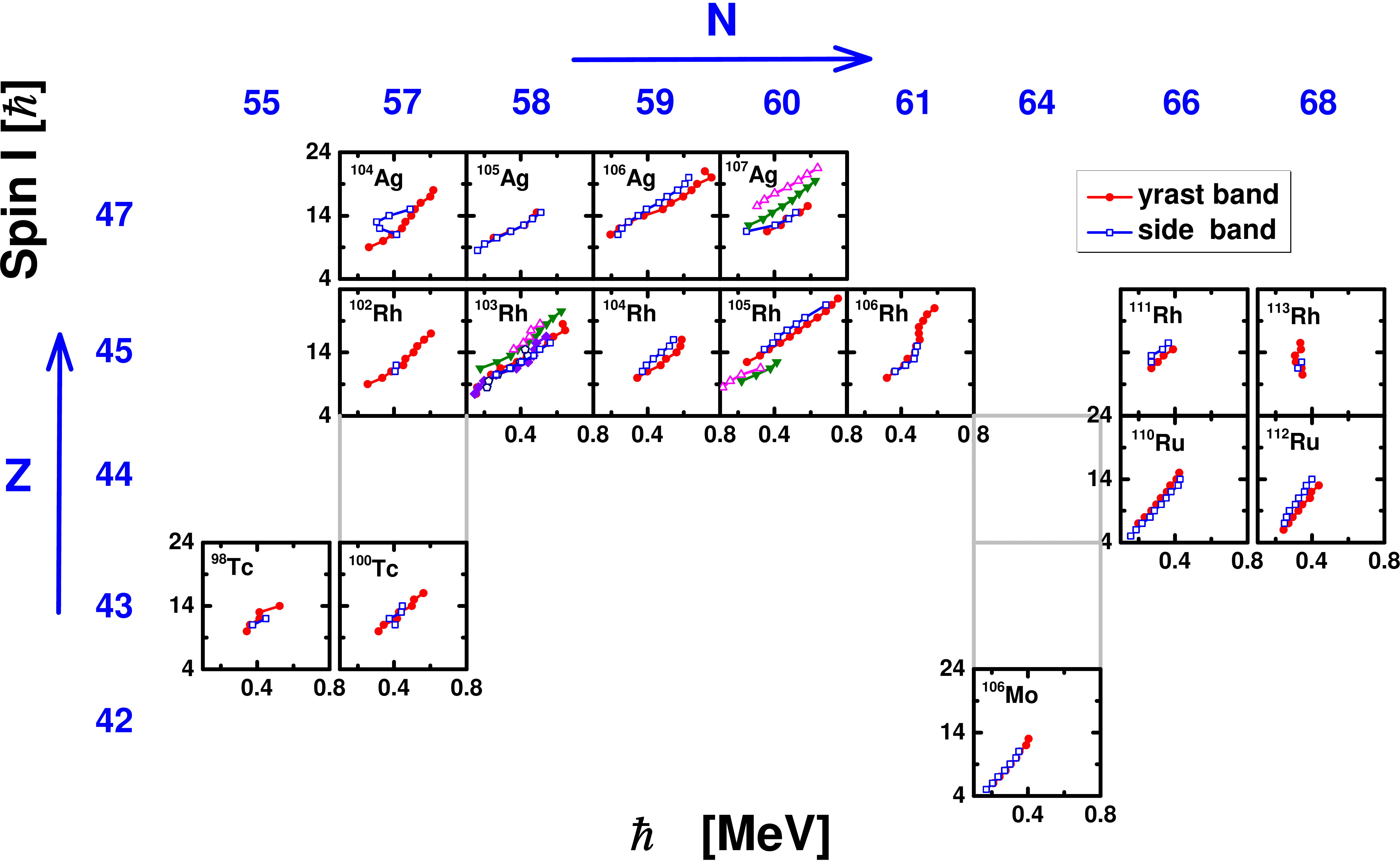}\\
  \caption{(Color online) The relation between the spin and the
   rotational frequency for chiral doublet bands in $A\sim$~100 mass
   region.}
  \label{fig16}
\end{figure}

\begin{figure}[htbp]
  \centering
  \includegraphics[height=15.5cm]{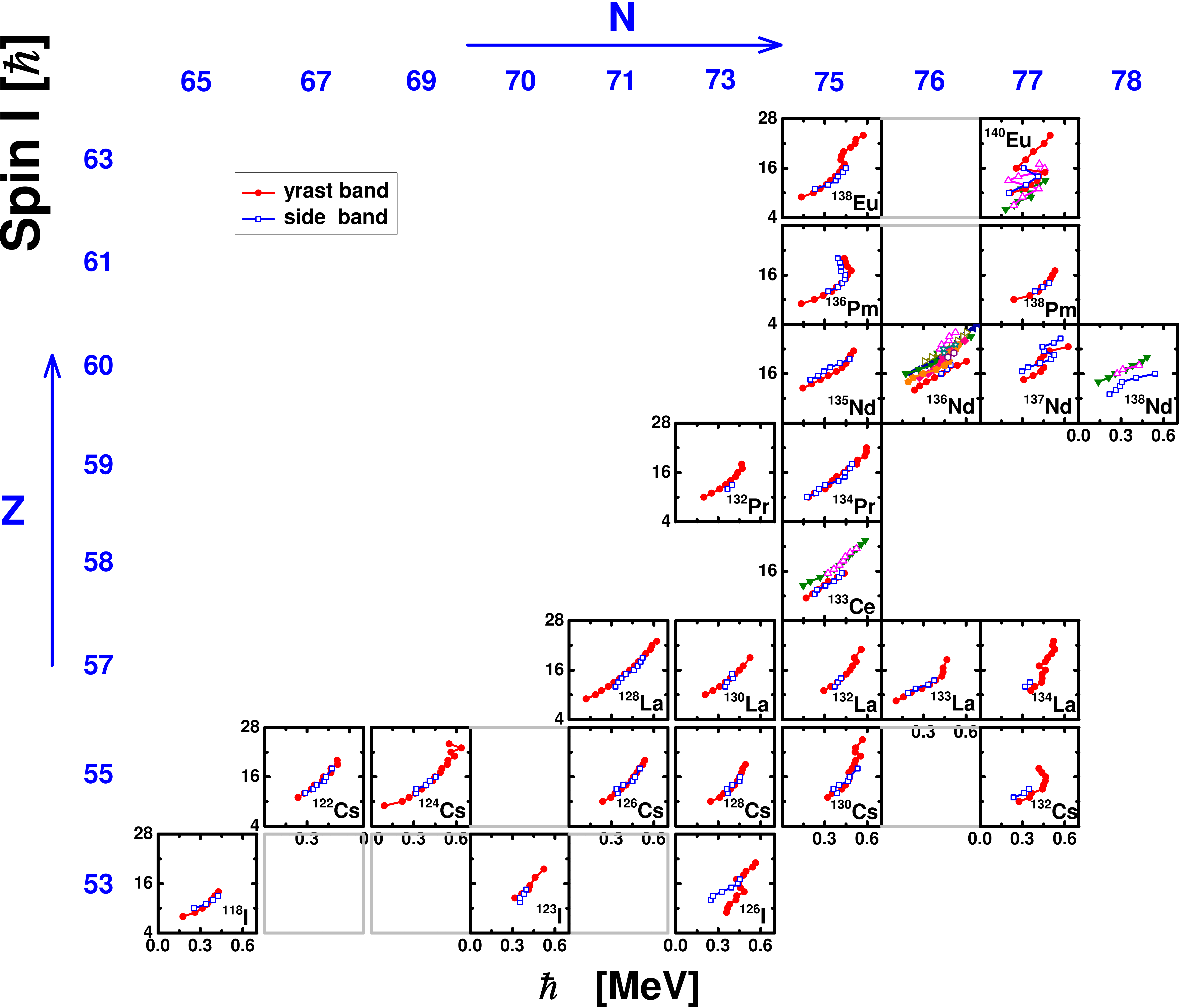}\\
  \caption{(Color online) The relation between the spin and the
   rotational frequency for chiral doublet bands in $A\sim$~130 mass
   region.}
  \label{fig17}
\end{figure}

\begin{figure}[htbp]
  \centering
  \includegraphics[height=9cm]{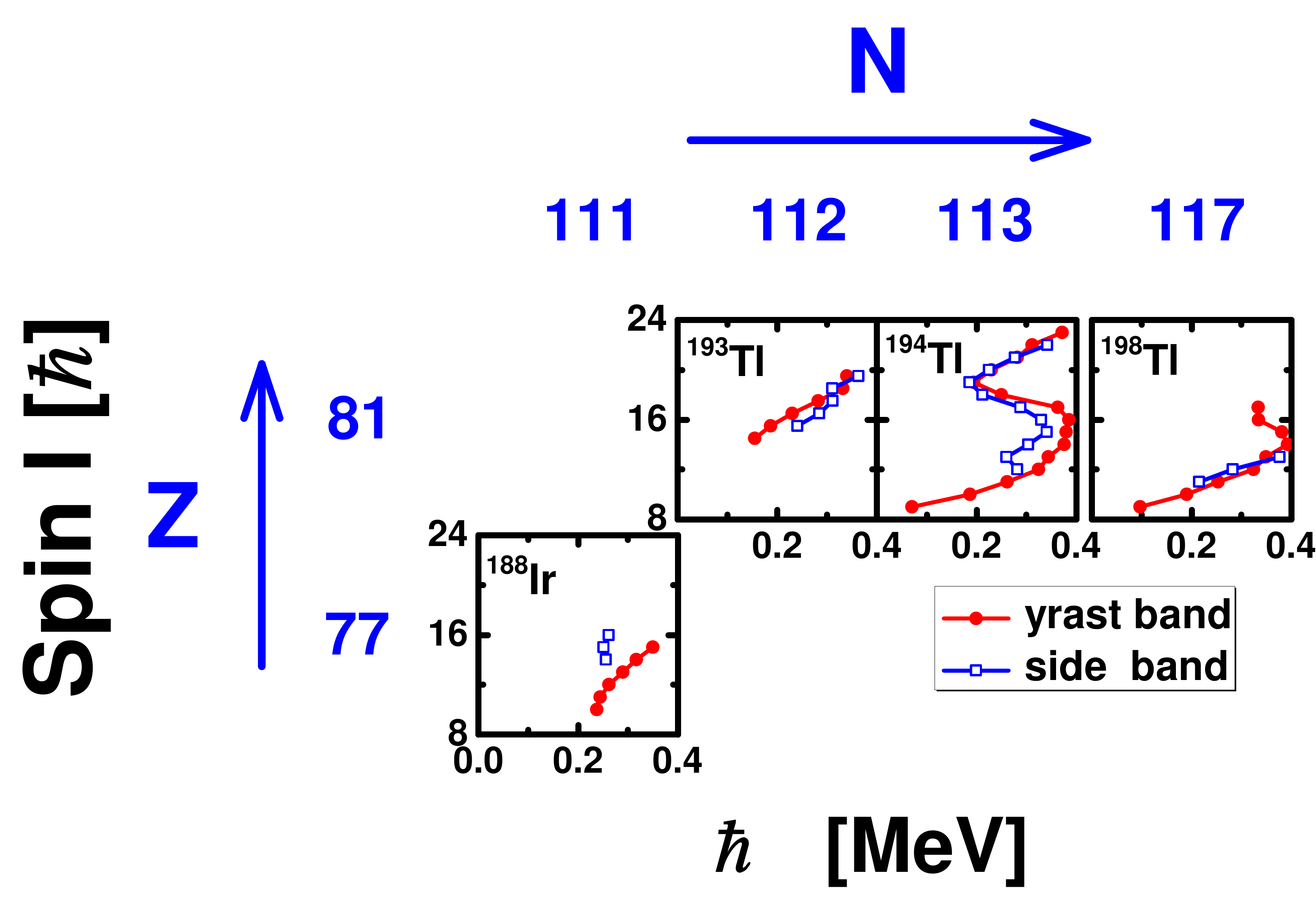}\\
  \caption{(Color online) The relation between the spin and the
   rotational frequency for chiral doublet bands in $A\sim$~190 mass
   region.}
  \label{fig18}
\end{figure}

In order to obtain the response of angular momentum alignments to rotation
frequencies and examine the similarities of the configurations for chiral
partner bands, the $I-\hbar\omega$ relation has been extracted. From the
rotational frequency $\hbar\omega$ defined as \cite{
Frauendorf1996Interpretation}, $\hbar\omega(I)=[E(I+1)-E(I-1)]/2$, the
relations between the spins and the rotational frequencies for all chiral
doublet bands in $A\sim$~80,~100,~130,~and 190 mass regions are shown in
Figs. \ref{fig15}-\ref{fig18}, respectively.

Generally, the $I-\hbar \omega$ relation for yrast band and side band is
similar, except for $^{104}$Ag, $^{107}$Ag, $^{126}$I, $^{136}$Nd,
$^{137}$Nd, $^{188}$Ir, and $^{193}$Tl. Possible backbending in some
nuclei exists.

\newpage

\newpage

\subsection{Kinematic moment of inertia}

\begin{figure}[htbp]
  \centering
  \includegraphics[height=5.8cm]{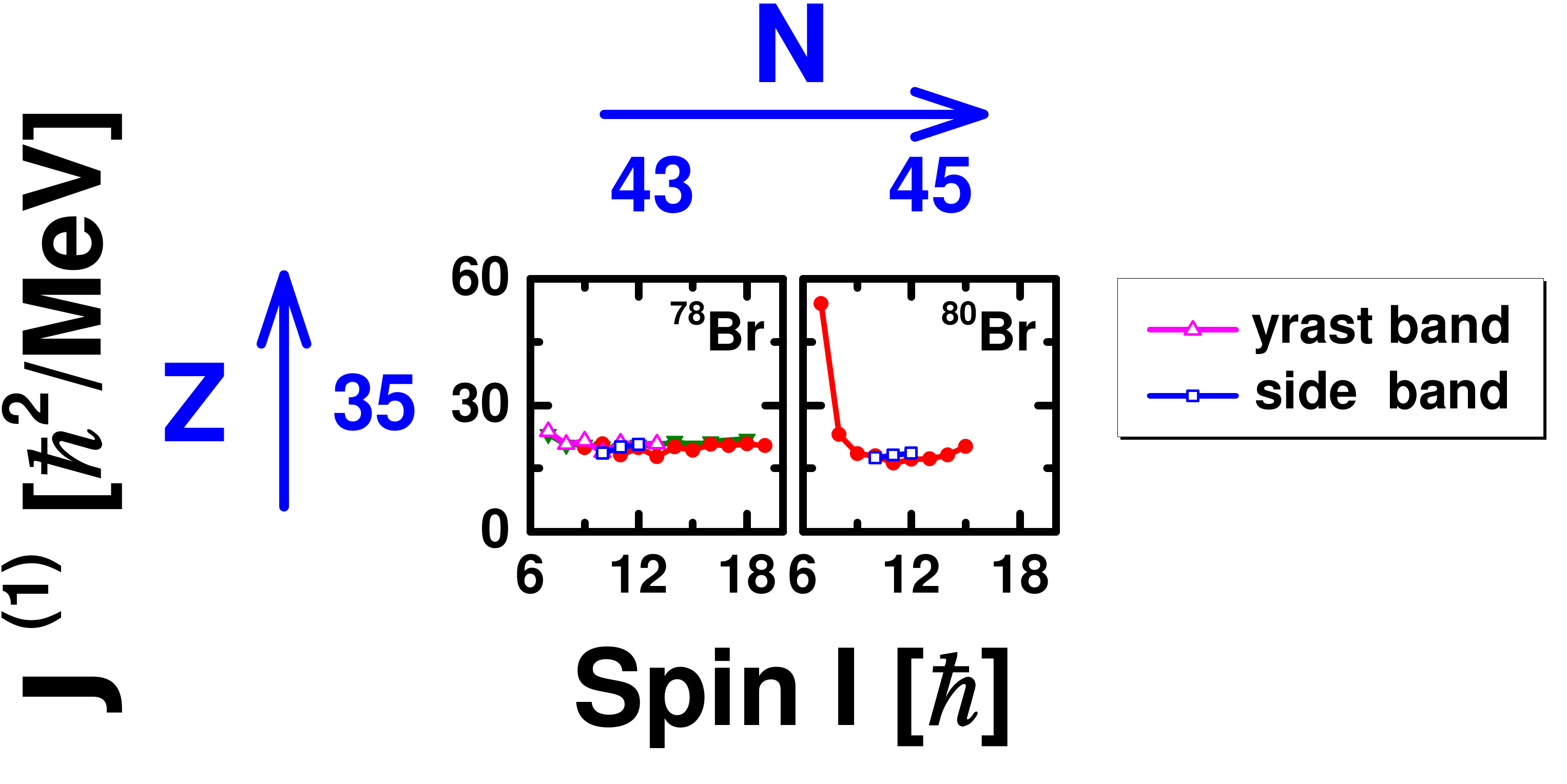}\\
  \caption{(Color online) Kinematic moments of inertia versus spin
    for chiral doublet bands in $A\sim$~80 mass region.}
  \label{fig19}
\end{figure}

\begin{figure}[htbp]
  \centering
  \includegraphics[height=11.5cm]{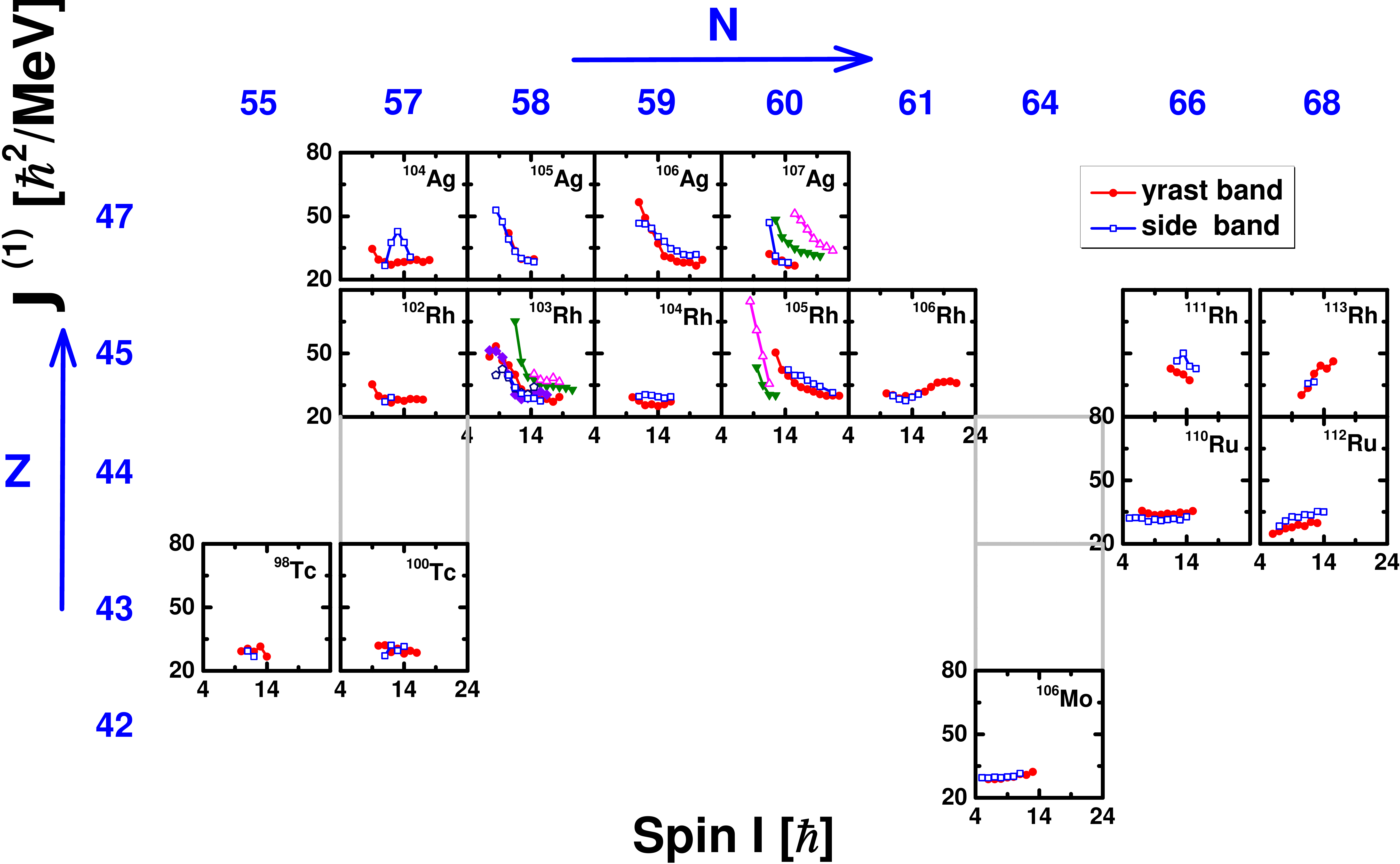}\\
  \caption{(Color online) Kinematic moments of inertia versus spin
    for chiral doublet bands in $A\sim$~100 mass region.}
  \label{fig20}
\end{figure}

\begin{figure}[htbp]
  \centering
  \includegraphics[height=15.5cm]{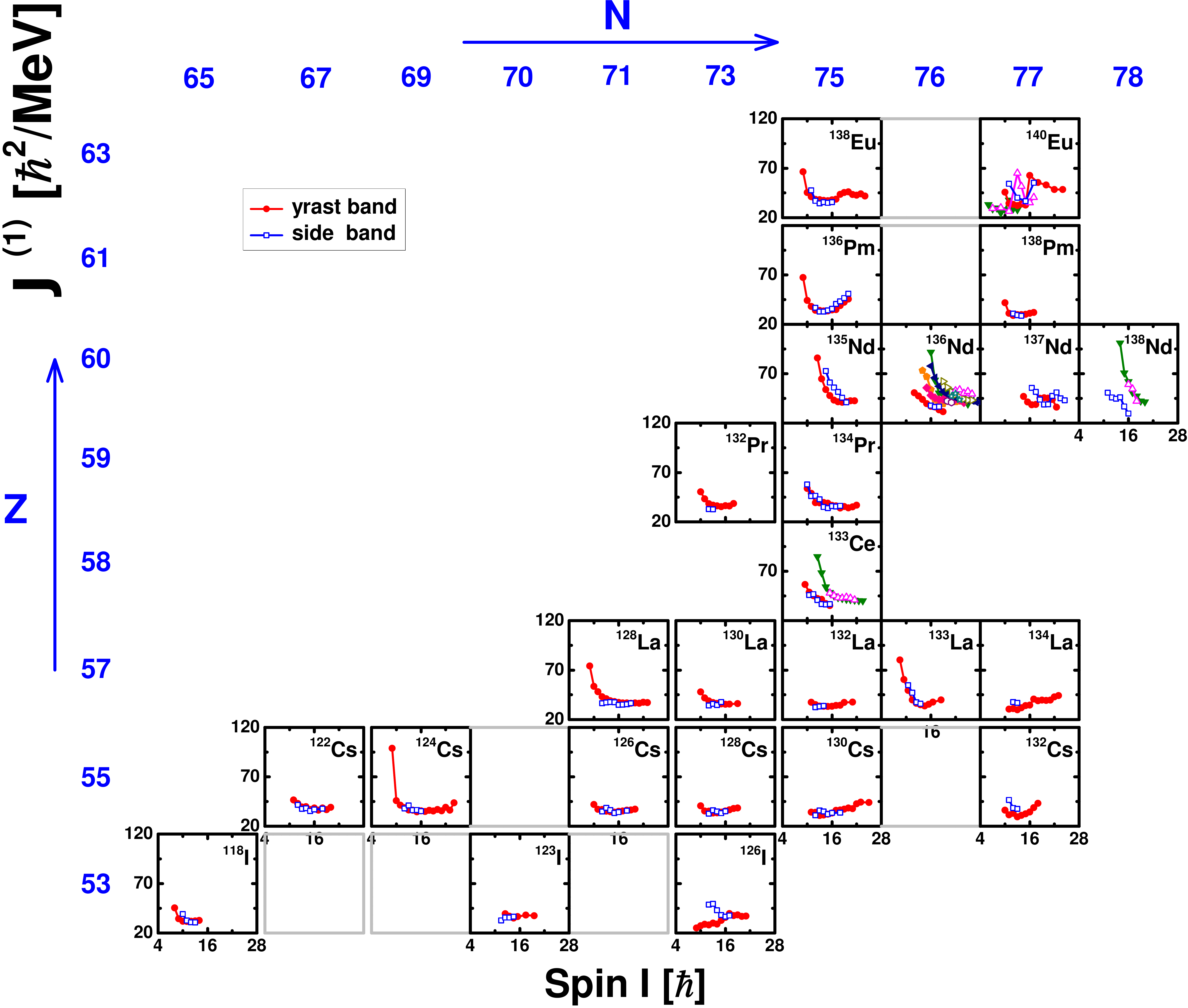}\\
  \caption{(Color online) Kinematic moments of inertia versus spin
    for chiral doublet bands in $A\sim$~130 mass region.}
  \label{fig21}
\end{figure}

\begin{figure}[htbp]
  \centering
  \includegraphics[height=9cm]{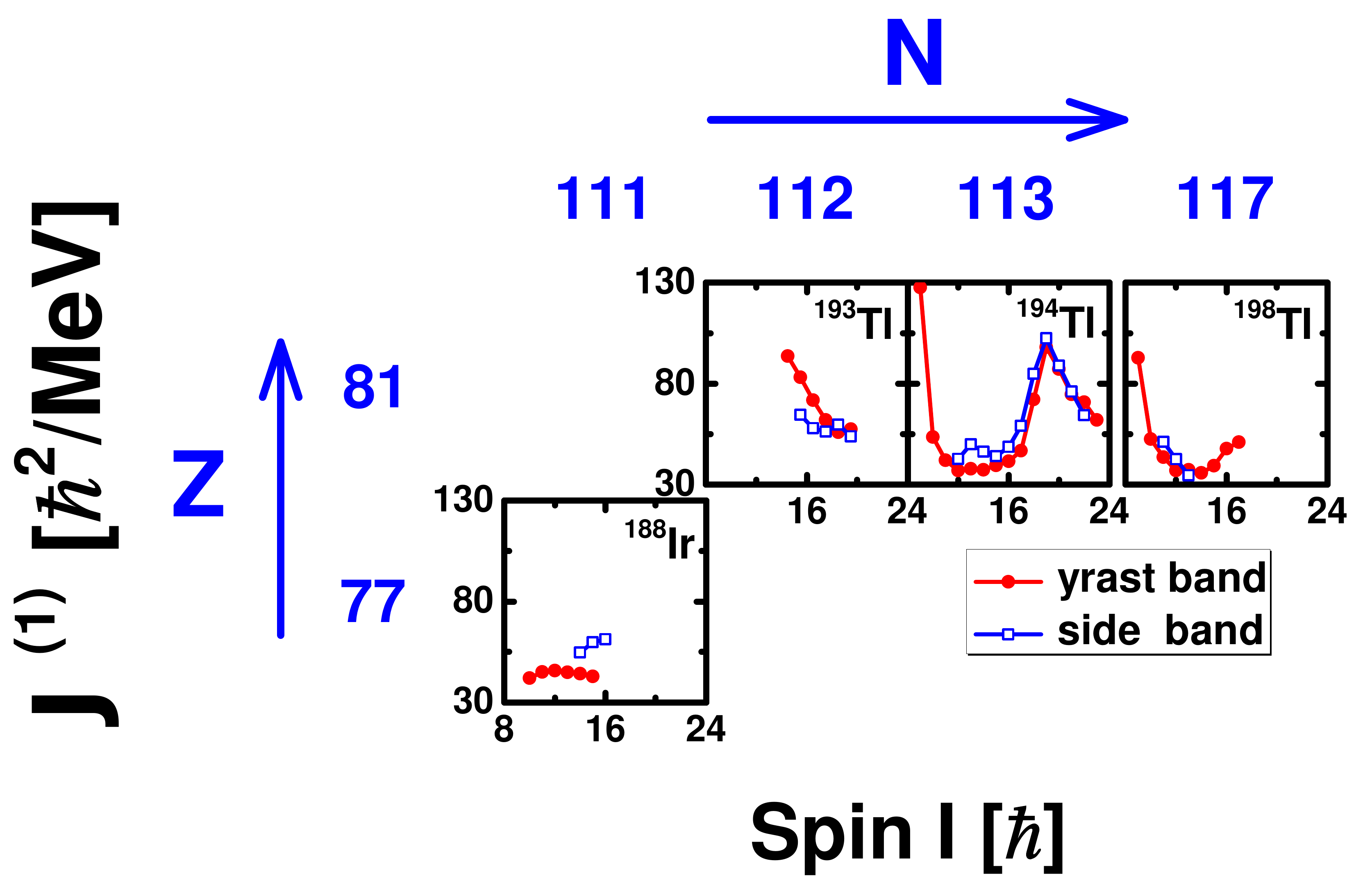}\\
  \caption{(Color online) Kinematic moments of inertia versus spin
    for chiral doublet bands in $A\sim$~190 mass region.}
  \label{fig22}
\end{figure}

From the definition $\mathcal{J}^{(1)}(I)=I/\hbar\omega(I)$, the
kinematic moments of inertia $\mathcal{J}^{(1)}$ for all chiral
doublet bands in $A\sim 80,~100,~130$, and 190 mass regions are
shown in Figs. \ref{fig19}-\ref{fig22}, respectively.

Generally, the kinematic moment of inertia for yrast band and side
band is similar, except for $^{104}$Ag, $^{107}$Ag, $^{126}$I,
$^{136}$Nd, $^{137}$Nd, $^{140}$Eu, and $^{188}$Ir. Normally, the
kinematic moment of inertia remains roughly constant.

\newpage

\newpage

\newpage

\subsection{Dynamic moment of inertia}

\begin{figure}[htbp]
  \centering
  \includegraphics[height=5.8cm]{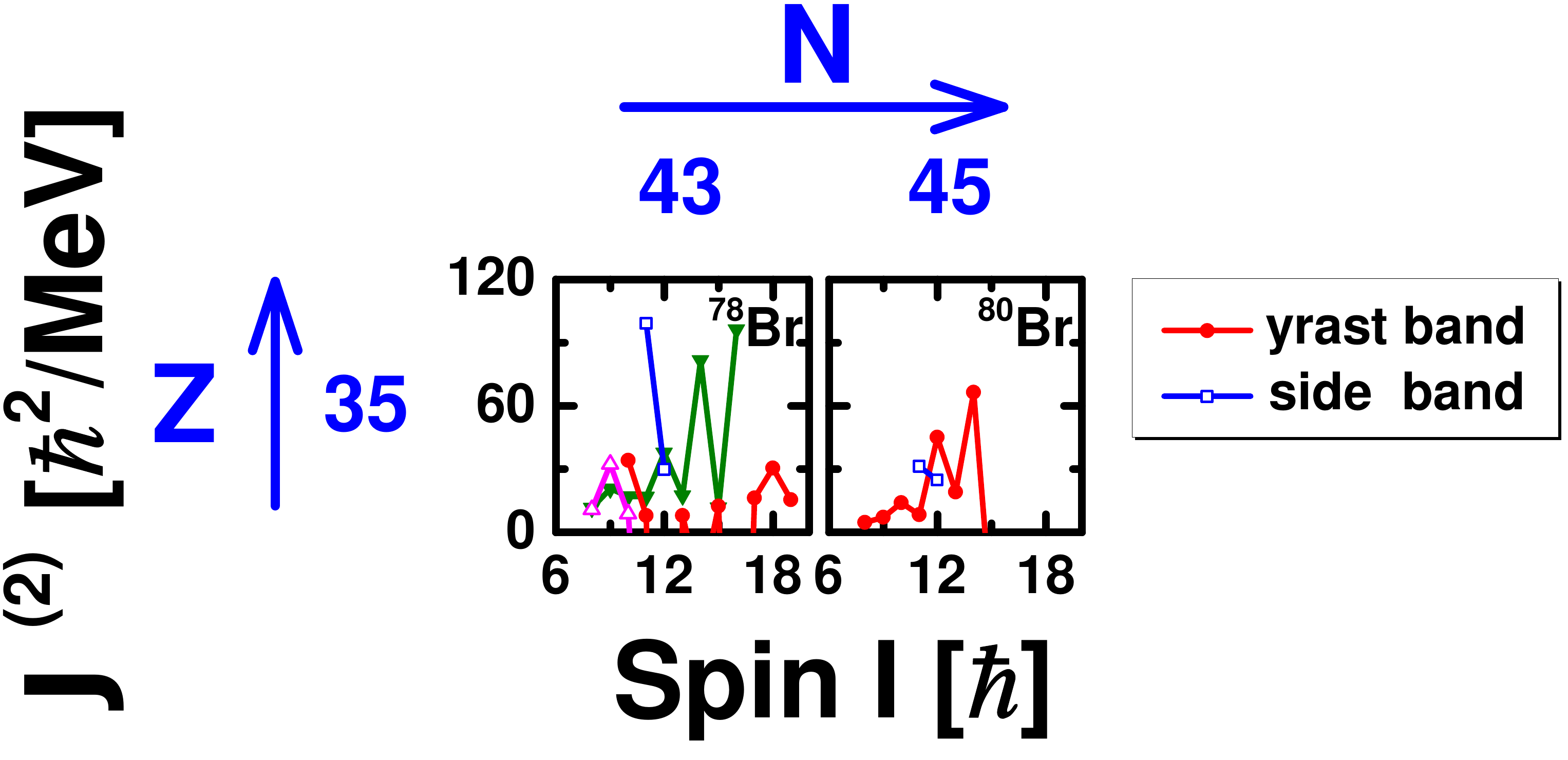}\\
  \caption{(Color online) Dynamic moments of inertia versus spin
    for chiral doublet bands in $A\sim$~80 mass region.}
  \label{fig23}
\end{figure}

\begin{figure}[htbp]
  \centering
  \includegraphics[height=11.5cm]{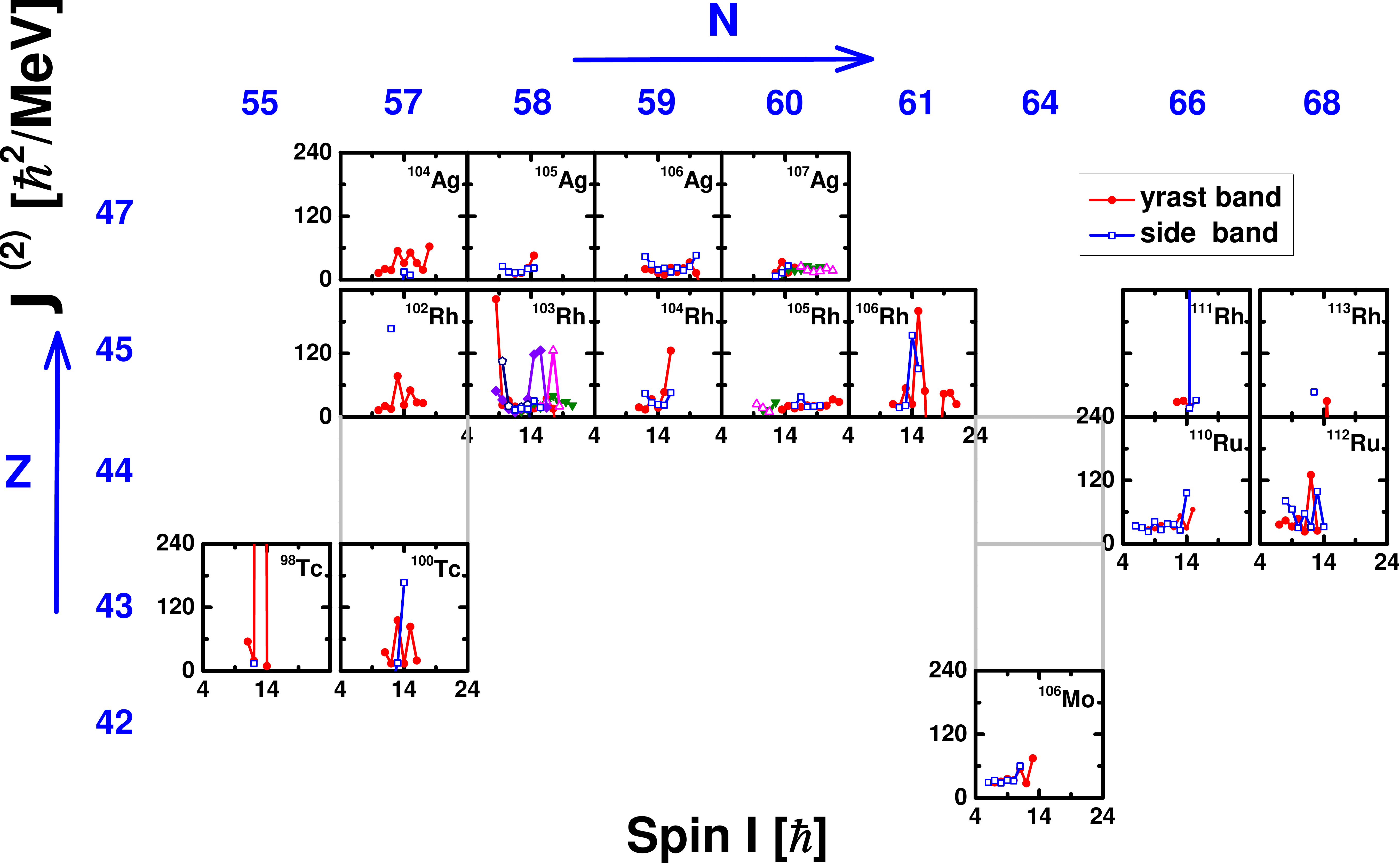}\\
  \caption{(Color online) Dynamic moments of inertia versus spin
    for chiral doublet bands in $A\sim$~100 mass region.}
  \label{fig24}
\end{figure}

\begin{figure}[htbp]
  \centering
  \includegraphics[height=15.5cm]{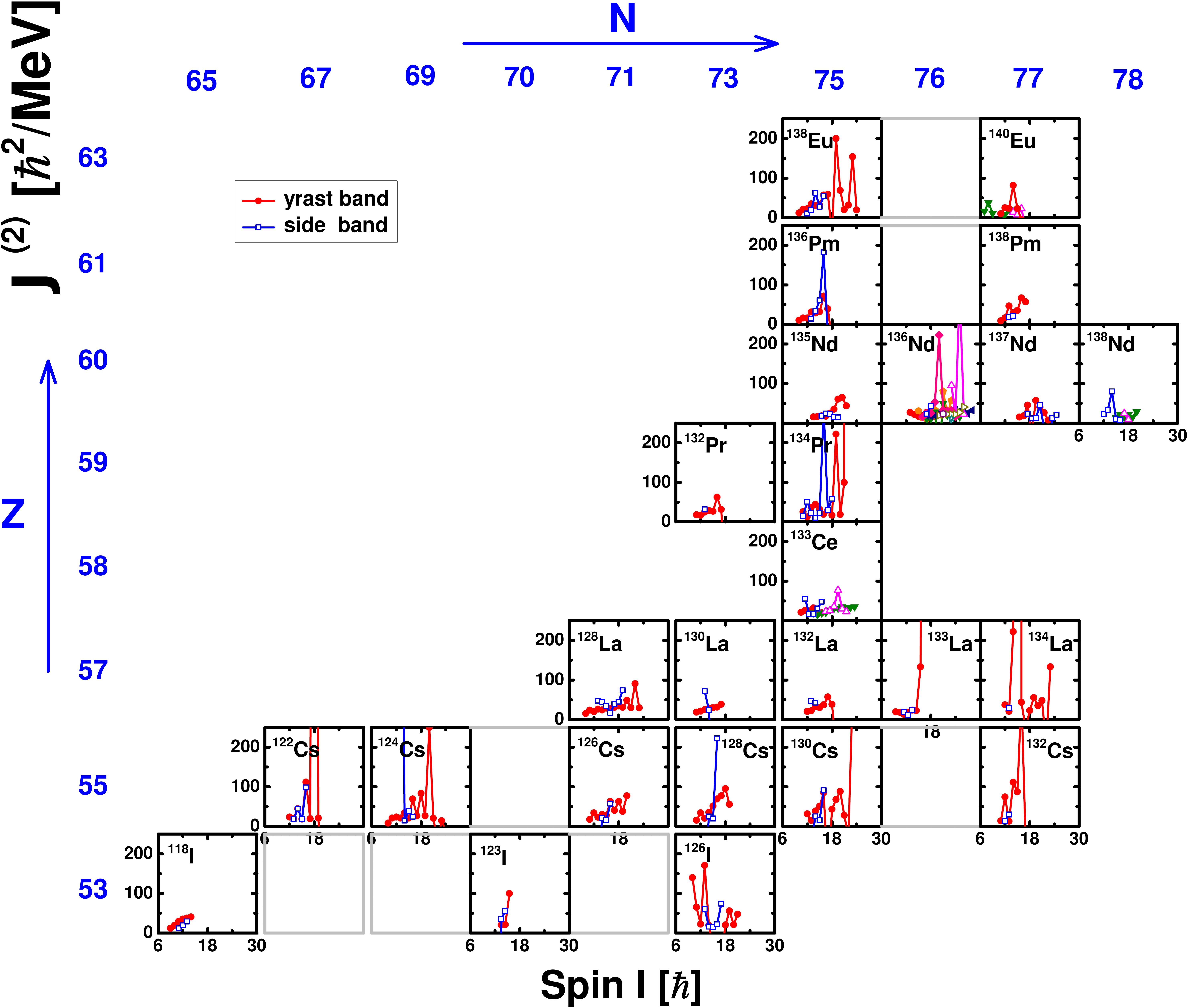}\\
  \caption{(Color online) Dynamic moments of inertia versus spin
    for chiral doublet bands in $A\sim$~130 mass region.}
  \label{fig25}
\end{figure}

\begin{figure}[htbp]
  \centering
  \includegraphics[height=9cm]{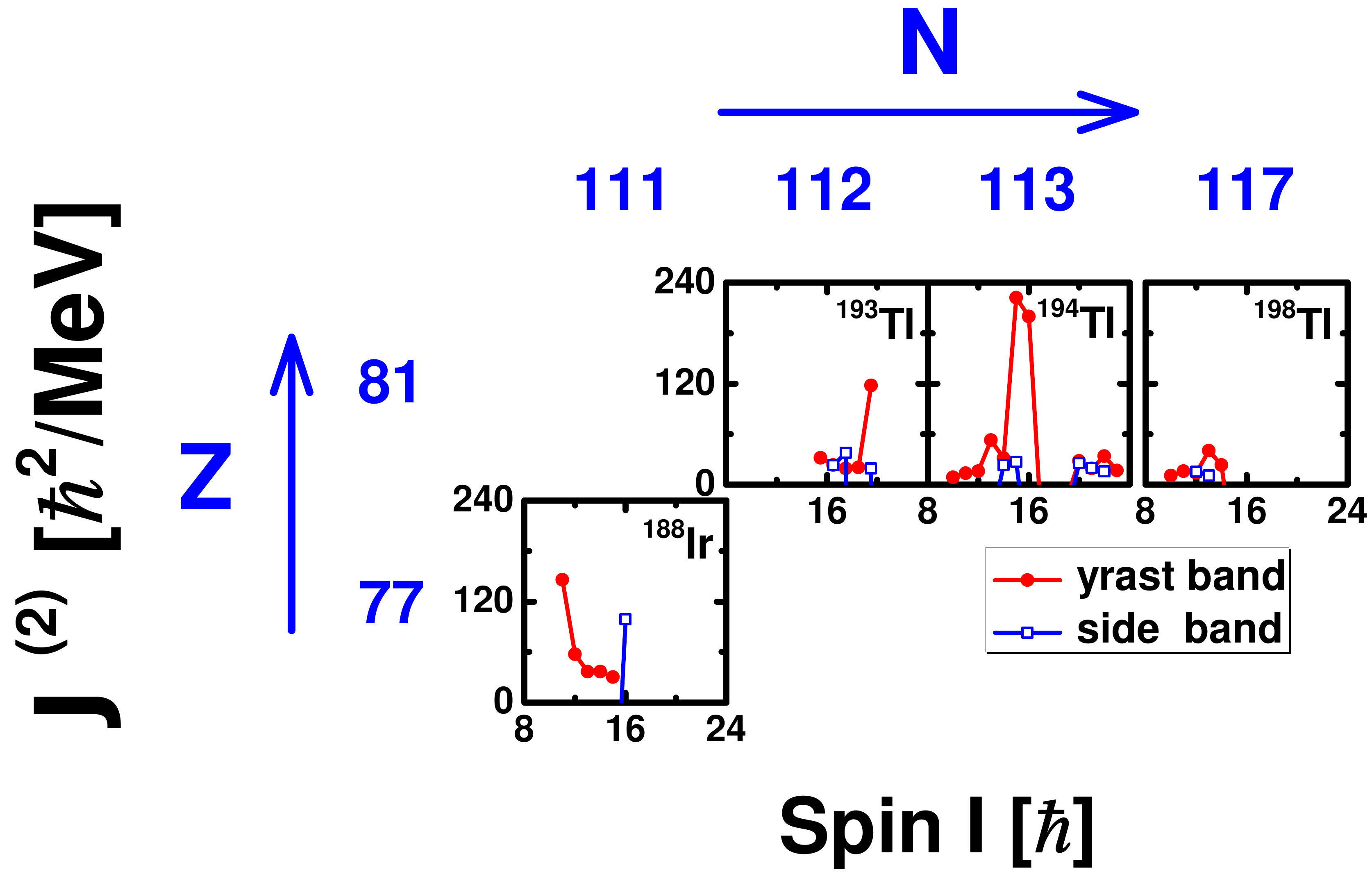}\\
  \caption{(Color online) Dynamic moments of inertia versus spin
    for chiral doublet bands in $A\sim$~190 mass region.}
  \label{fig26}
\end{figure}

From the definition $\mathcal{J}^{(2)}(I)=\hbar~\mathrm{d}I/
\mathrm{d}\omega(I)$, the dynamic moments of inertia $\mathcal{J}^
{(2)}$ for all chiral doublet bands in $A\sim 80,~100,~130$, and
190 mass regions are shown in Figs. \ref{fig23}-\ref{fig26},
respectively.

Since the $\mathcal{J}^{(2)}$ corresponds to the second derivative
of the energy with the spin, a large fluctuation exists.
Nevertheless, similarities between the yrast bands and the side
bands still exist in most chiral doublet bands.

\newpage

\newpage

\newpage

\newpage

\subsection{Electromagnetic transition probability}

\begin{figure}[htbp]
  \centering
  \includegraphics[height=7.4cm]{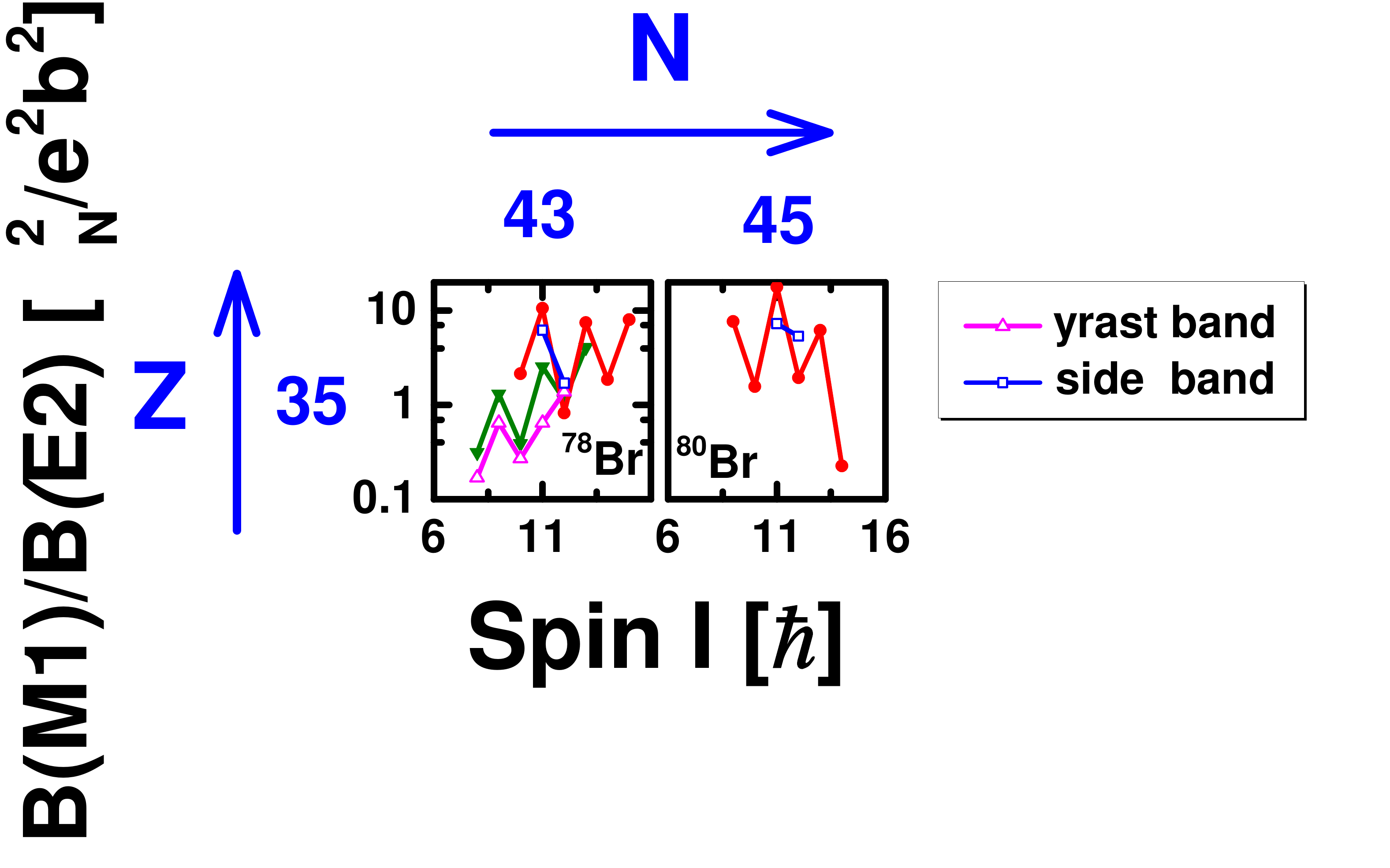}\\
  \caption{(Color online) $B(M1)/B(E2)$~ratios versus spin for
    chiral doublet bands in the $A\sim$~80 mass region.}
  \label{fig27}
\end{figure}

\begin{figure}[htbp]
  \centering
  \includegraphics[height=11.5cm]{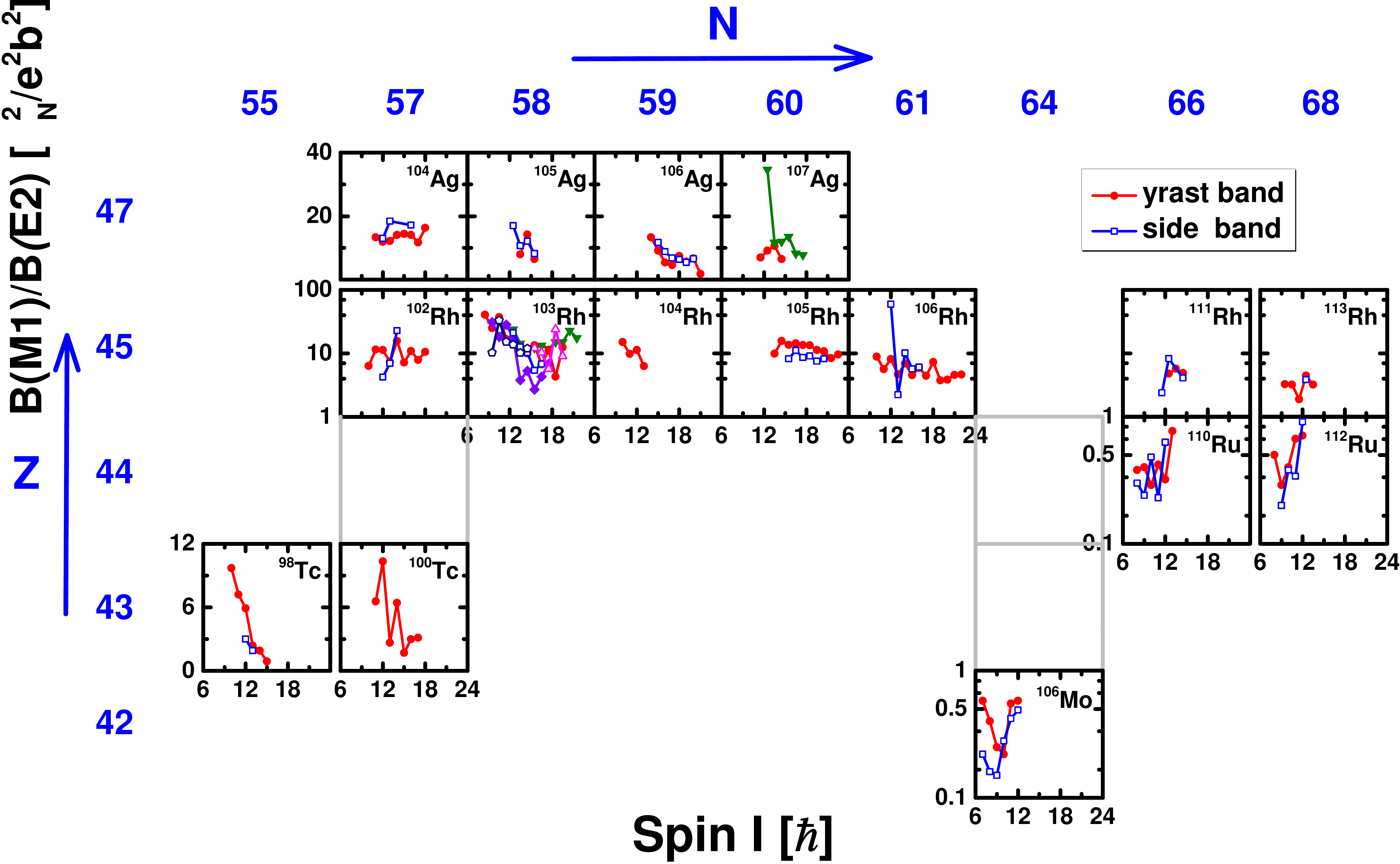}\\
  \caption{(Color online) $B(M1)/B(E2)$~ratios versus spin for
    chiral doublet bands in the $A\sim$~100 mass region.}
  \label{fig28}
\end{figure}

\begin{figure}[htbp]
  \centering
  \includegraphics[height=15.5cm]{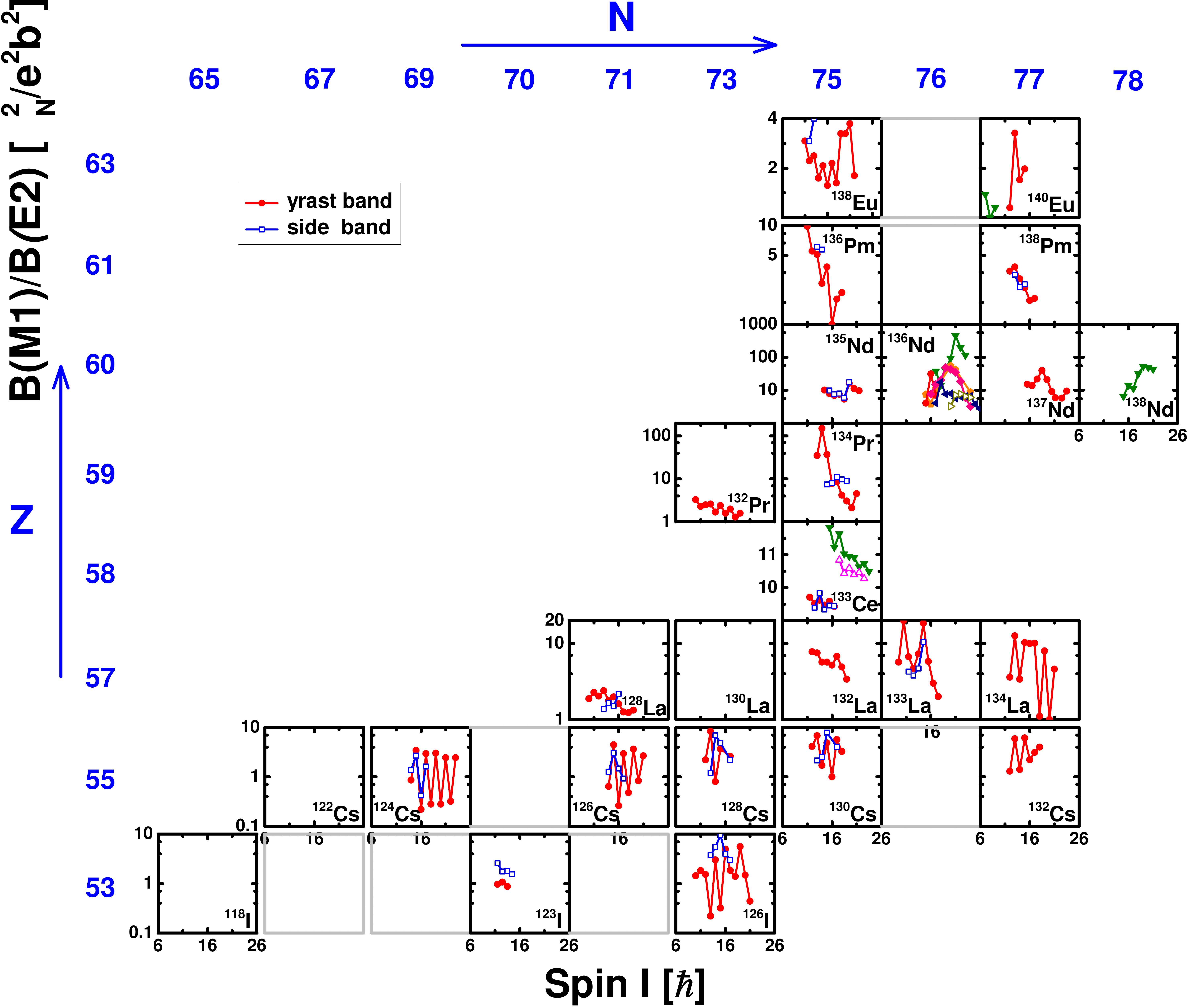}\\
  \caption{(Color online) $B(M1)/B(E2)$~ratios versus spin for
    chiral doublet bands in the $A\sim$~130 mass region.}
  \label{fig29}
\end{figure}

\begin{figure}[htbp]
  \centering
  \includegraphics[height=9cm]{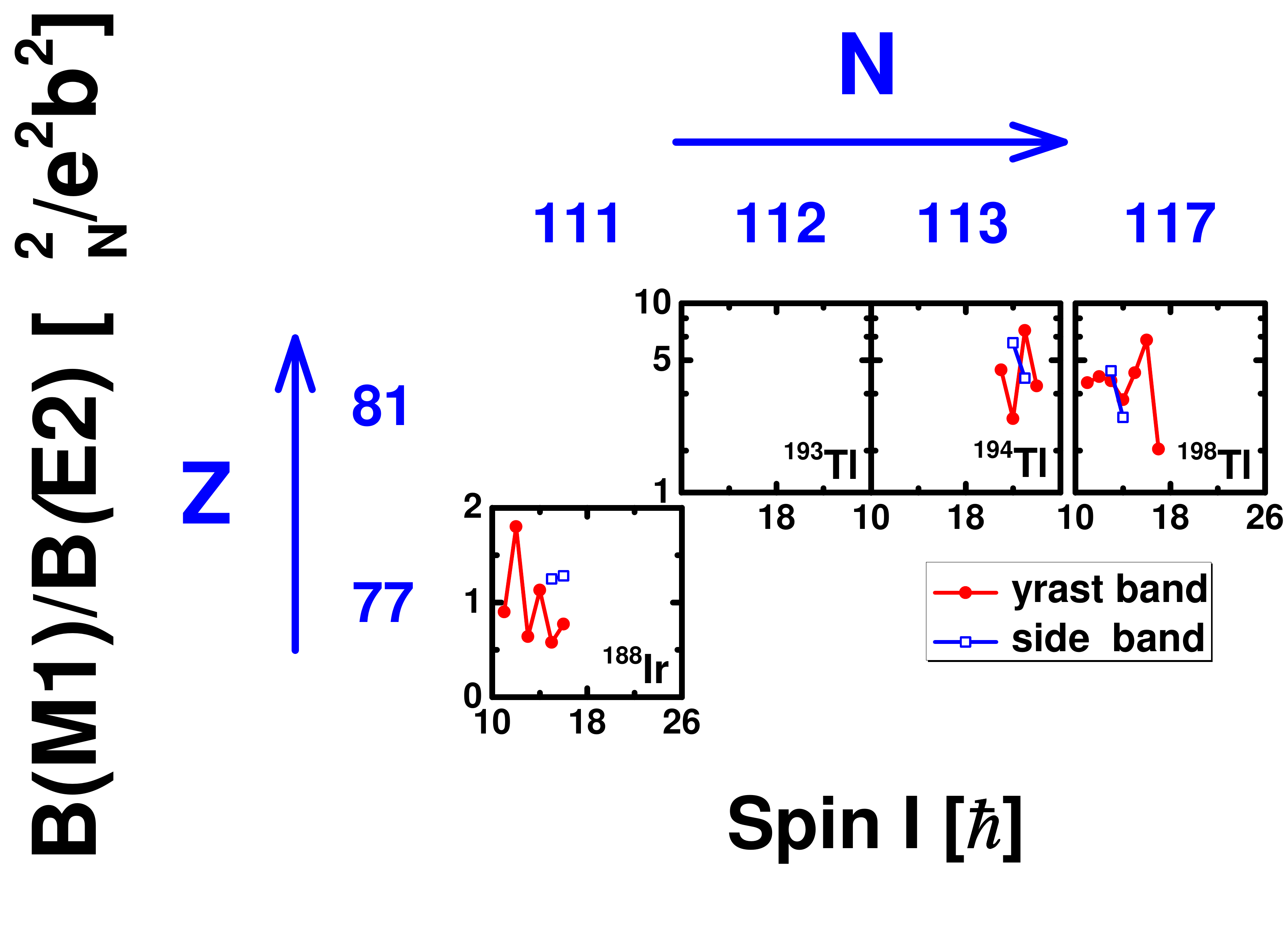}\\
  \caption{(Color online) $B(M1)/B(E2)$~ratios versus spin for
    chiral doublet bands in the $A\sim$~190 mass region.}
  \label{fig30}
\end{figure}

The ratios of the magnetic dipole transition strength to the
electric quadrupole transition strength $B(M1)/B(E2)$ for all
chiral doublet bands in $A\sim$~80,~100,~130,~and 190 mass regions
are given in Figs. \ref{fig27}-\ref{fig30}, respectively. For the
ideal chiral bands, due to the restoration of the chiral symmetry
in the laboratory frame there are phase consequences for the
chiral wavefunctions resulting in $M$1 and $E$2 selection rules
which can manifest as $B(M1)/B(E2)$ ratios staggering as a function
of spin. And the $B(M1)/B(E2)$ ratios are expected to be very
similar for the chiral partner bands \cite{Wang2007Examining}.

For the nuclei $^{106}$Mo, $^{98}$Tc, $^{100}$Tc, $^{110}$Ru,
$^{112}$Ru, $^{102}$Rh, $^{106}$Rh, $^{111}$Rh, $^{113}$Rh,
$^{104}$Ag, $^{126}$I, $^{128}$Cs, $^{132}$Cs, $^{132}$La,
$^{134}$La, $^{134}$Pr, $^{136}$Nd, $^{137}$Nd, $^{138}$Nd,
$^{188}$Ir, and $^{198}$Tl, the $B(M1)/B(E2)$ ratios are extracted
by the equation (2) in Ref. \cite{Zhang1999High}. For the nuclei
$^{104}$Rh, $^{106}$Ag, $^{107}$Ag, $^{124}$Cs, $^{126}$Cs,
$^{130}$Cs, $^{135}$Nd, and $^{194}$Tl with $B(M1)$ and $B(E2)$
values available, the $B(M1)/B(E2)$ ratios are also calculated. The
other data available are from the original references. The
staggering of the $B(M1)/B(E2)$ ratios exists in most chiral doublet
bands.

\newpage

\newpage

\newpage

\newpage

\section{Summary}

Since the prediction of nuclear chirality in 1997, the nuclear
chirality has become one of the hot topics in current nuclear
physics frontiers. Experimentally, 59 chiral doublet bands in
47 chiral nuclei (including 8 nuclei with multiple chiral
doublets) have been reported in $A\sim80,~100,~130$, and~190
mass regions.

The spins, parities, energies, ratios of the magnetic dipole
transition strengths to the electric quadrupole transition
strengths, and related references for these nuclei have been
compiled and listed in Table 1. For these nuclei with the
magnetic dipole transition strengths and the electric
quadrupole transition strengths measured, the corresponding
results have been given in Table 2. A brief discussion has been
provided after the presentation of energy $E$, energy difference
$\Delta E$, energy staggering parameter $S(I)$, rotational
frequency $\omega$, kinematic moment of inertia
$\mathcal{J}^{(1)}$, dynamic moment of inertia $\mathcal{J}^
{(2)}$, and ratio of the magnetic dipole transition strength to
the electric quadrupole transition strength $B(M1)/B(E2)$ versus
spin $I$ in each mass region.

\ack
The authors are indebted to Prof. J. Meng for the suggestion of
this topic and the guidance during this work. The authors express
thanks to R. Bark, U. Garg, A. A. Hecht, P. Joshi, T. Koike, C. Liu,
M. L. Liu, J. B. Lu, Y. X. Luo, K. Y. Ma, Y. J. Ma, G. Rainovski,
A. K. Singh, K. Starosta, J. Tim$\mathrm{\acute{a}}$r, B. Wadsworth,
S. Y. Wang, Y. Zheng, L. H. Zhu, S. F. Zhu, and S. J. Zhu for
providing the data and helpful suggestions. Fruitful discussions
with F. Q. Chen, Q. B. Chen, Z. Shi, Y. K. Wang, X. H. Wu,
S. Q. Zhang, Z. H. Zhang, and P. W. Zhao are very much appreciated.
This work is supported in part by the Major State 973 Program of
China (Grant No. 2013CB834400) and the National Natural Science
Foundation of China (Grants No. 11335002, No. 11375015, No.
11461141002, and No. 11621131001).

\clearpage

\TableExplanation

\section*{Table 1.\label{tbl1te} Chiral doublet bands.}

\begin{tabular*}{0.95\textwidth}{@{}@{\extracolsep{\fill}}lp{5.5in}@{}}

  $^A_Z\!X_N$	
            &Denotes the specific nuclide with\\
            &~~~~~$X$ chemical symbol\\
            &~~~~~$A$ mass number\\
            &~~~~~$Z$ atomic number\\
            &~~~~~$N$ neutron number\\
            & A horizontal line connecting a row marks the end of
            entry for each chiral nucleus. For multiple chiral
            doublet bands, the pairs of chiral doublet bands are
            separated by a horizontal line starting from the second
            column.\\

  $I^\pi$
	        & $I$ denotes the level spin in units of $\hbar$ for each
            band member. $\pi$ denotes the parity ($+$ or $-$).
            References for each nucleus are given directly in last
            column.\\

  $E$
	        & Level energy in units of keV. The number of digits
            follows the original experimental article. In the energy
            column, yrast and side denote the two partners of chiral
            doublet bands. The star marks the level which is the
            yrast state (the lowest state for given spin) in the
            nucleus.\\
\end{tabular*}

\section*{Table 2.\label{tbl2te} Chiral doublet bands with $B(M1)$ and
$B(E2)$ values.}

\begin{tabular*}{0.95\textwidth}{@{}@{\extracolsep{\fill}}lp{5.5in}@{}}

  $^A_Z\!X_N$
            & Same as the explanation in Table 1.\\

  $I^\pi$
            & Same as the explanation in Table 1.\\

  $E$
            & Same as the explanation in Table 1.\\

  $B(M1)$	
            & The magnetic dipole transition strength in units of
            $\mu^2_N$ with the uncertainties available in parentheses.
            The number of digits follows the original experimental
            article. In the $B(M1)$ column, yrast and side denote the
            two partners of chiral doublet bands.\\

  $B(E2)$
	        & The electric quadrupole transition strength in units of
            $\mathrm{e}^2\mathrm{b}^2$ with the uncertainties
            available in parentheses. The number of digits follows the
            original experimental article. In the $B(E2)$ column, yrast
            and side denote the two partners of chiral doublet bands.\\
\end{tabular*}
\label{tableII}

\renewcommand{\arraystretch}{1.0}

\bigskip

\newpage

\datatables 


\setlength{\LTleft}{0pt}
\setlength{\LTright}{0pt}


\setlength{\tabcolsep}{0.5\tabcolsep}

\renewcommand{\arraystretch}{1.0}


\begin{longtable}{@{\extracolsep\fill}ccccccc}
    \caption{Chiral doublet bands. See page\ \pageref{tbl1te} for
      Explanation of Tables for details.}

      \multirow{2}{*}{\centering Nuclei}        &
      \multirow{2}{*}{\centering$I^\pi(\hbar)$} &
      \multicolumn{2}{c}{\centering $E$~(keV)}  &
      \multicolumn{2}{c}{$B(M1)/B(E2)~
       (\mu^2_N/\mathrm{e}^2\mathrm{b}^2)$}     &
      \multirow{2}{*}{\centering References}\\

      \cline{3-4}\cline{5-6} &  & yrast & side  & yrast & side &\\

      \hline
\endfirsthead

      \multicolumn{3}{@{}c@{}}{Table 1 (continued)}\\
      \hline

      \multirow{2}{*}{\centering Nuclei}        &
      \multirow{2}{*}{\centering$I^\pi(\hbar)$} &
      \multicolumn{2}{c}{\centering $E$~(keV)}  &
      \multicolumn{2}{c}{$B(M1)/B(E2)~
       (\mu^2_N/\mathrm{e}^2\mathrm{b}^2)$}     &
      \multirow{2}{*}{\centering References}\\
      \cline{3-4}\cline{5-6} &  & yrast & side  & yrast & side &\\

      \hline
\endhead

      \hline
      \noalign{\vskip3pt}
      \multicolumn{7}{r}{\itshape Continued on next page}\\
\endfoot

      \noalign{\vskip3pt}\hline
\endlastfoot

   $^{78}_{35}\!\mathrm{Br}_{43}$ & $8^+$ & $465.7^*$  & & & &
   \cite{Liu2016Evidence,Liu2018Br}\\

   &  $9^+ $  &  $975.3^* $   &  1408.6  &      &      &    \\
   &  $10^+$  &  $1369.4^*$   &  1881.5  & 2.13918
     ($^{+0.07423}_{-0.07423}$)                 &      &    \\
   &  $11^+$  &  $1937.8^*$   &  2482.8  & 10.56231
     ($^{+0.92806}_{-0.92806}$)          &
     6.15398~($^{+0.88693}_{-0.88693}$)                &    \\
   &  $12^+$  &  $2582.5^*$   &  2975.9  &$< 0.82258$  &
     1.6983~($^{+0.40338}_{-0.40338}$)                 &    \\
   &  $13^+$  &  $3146.0^*$   &  3644.4  &
     7.4299~($^{+0.6144}_{-0.6144}$)            &      &    \\
   &  $14^+$  &  $4046.2^*$   &          &$< 1.85685$  &
                                                       &    \\
   &  $15^+$  &  $4539.0^*$   &          &
     8.02873~($^{+1.56134}_{-1.56134}$)         &      &    \\
\end{longtable}

\newpage

\begin{longtable}{@{\extracolsep\fill}ccccccccc}

 \caption{Chiral doublet bands with $B(M1)$ and $B(E2)$ values.
 See page\ \pageref{tbl2te} for Explanation of Tables for details.}
  \multirow{2}{*}{\centering Nuclei} &\multirow{2}{*}{\centering $I^\pi~
  (\hbar)$}& \multicolumn{2}{c}{\centering $E$~(keV)} &\multicolumn{2}{c}
  {\centering $B(M1)~(\mu^2_N)$}  &\multicolumn{2}{c}{\centering $B(E2)~
  (\mathrm{e}^2\mathrm{b}^2)$}&\multirow{2}{*}{\centering References}\\

   \cline{3-4}  \cline{5-6}  \cline{7-8}

   & & yrast & side & yrast & side & yrast & side &      \\

   \hline

\endfirsthead

   \multicolumn{3}{@{}c@{}}{Table 2 (continued)}\\

   \hline

   \multirow{2}{*}{\centering Nuclei} &\multirow{2}{*}{\centering $I^\pi~
   (\hbar)$}& \multicolumn{2}{c}{\centering $E$~(keV)} &\multicolumn{2}{c}
   {\centering $B(M1)~(\mu^2_N)$}  &\multicolumn{2}{c}{\centering $B(E2)~
   (\mathrm{e}^2\mathrm{b}^2)$}&\multirow{2}{*}{\centering References}\\

   \cline{3-4}   \cline{5-6}  \cline{7-8}

   & & yrast & side & yrast & side & yrast & side & \\

   \hline

\endhead

   \hline
   \noalign{\vskip3pt}
   \multicolumn{9}{r}{\itshape Continued on next page}\\

\endfoot

   \noalign{\vskip3pt}\hline

\endlastfoot


        $^{103}_{~45}\!\mathrm{Rh}_{58}$     & $10.5^+$ &  3238
        &     &    &     &     &     &
        \cite{Timar2006Role, Kuti2014Multiple,Suzuki2008Lifetime}\\

        & $11.5^+$ & 3357     &      &    &   &   &   &      \\
        & $12.5^+$ & 3591     &      &    &   &   &   &      \\
        & $13.5^+$ & 3899     & 4445 &    &   &   &   &      \\
        & $14.5^+$ & 4281     & 4789 &    &   &   &   &      \\
        & $15.5^+$ & 4665     & 5166 &    &   &   &   &      \\
        & $16.5^+$ & 5156     & 5616 &    &   &   &   &      \\
        & $17.5^+$ & 5622     & 6062 &    &   &   &   &      \\
\end{longtable}


\section*{References}

\begin{thebibliography}{100}

\bibitem{Meng2016Relativistic}
J.~Meng (Ed.),
\newblock Relativistic Density Functional for Nuclear Structure,
International Review of Nuclear Physics,
\newblock vol.~10, World Scientific, Singapore, 2016, pp. 387-411.

\bibitem{Frauendorf1997Tilted}
S.~Frauendorf, J.~Meng,
\newblock Nuclear Phys. A, 617 (1997) 131.

\bibitem{Meng2010Open}
J.~Meng, S.~Q. Zhang,
\newblock J. Phys. G: Nucl. Part. Phys. 37 (2010) 064025.

\bibitem{Frauendorf2001Spontaneous}
S.~Frauendorf,
\newblock Rev. Mod. Phys. 73 (2001) 463.

\bibitem{Meng2011Chirality}
J.~Meng,
\newblock Int. J. Mod. Phys. E, 20 (2011) 341.

\bibitem{Meng2016Nuclear}
J.~Meng, P.~W. Zhao,
\newblock Phys. Scr. 91 (2016) 053008.

\bibitem{Zhou2016Multidimensionally}
S.~G. Zhou,
\newblock Phys. Scr. 91 (2016) 063008.

\bibitem{Sheikh2016Microscopic}
J.~A. Sheikh, G.~H. Bhat, W.~A. Dar, S.~Jehangir,
P.~A. Ganai,
\newblock Phys. Scr. 91 (2016) 063015.

\bibitem{RADUTA2016Specific}
A.~A. Raduta,
\newblock Prog. Part. Nucl. Phys. 90 (2016) 241.

\bibitem{Starosta2017Nuclear}
K.~Starosta, T.~Koike,
\newblock Phys. Scr. 92 (2017) 093002.

\bibitem{Peng2003Description}
J.~Peng, J.~Meng, S.~Q. Zhang,
\newblock Phys. Rev. C 68 (2003) 044324.

\bibitem{Koike2004Chiral}
T.~Koike, K.~Starosta, I.~Hamamoto,
\newblock Phys. Rev. Lett. 93 (2004) 172502.

\bibitem{Qi2009Examining}
B.~Qi, S.~Q. Zhang, S.~Y. Wang, J.~M. Yao, J.~Meng,
\newblock Phys. Rev. C 79 (2009) 041302(R).

\bibitem{Qi2010Band}
B.~Qi, S.~Q. Zhang, S.~Y. Wang, J.~Meng,
\newblock Chin. Phys. Lett. 27 (2010) 112101.

\bibitem{Chen2010Chiral}
Q.~B. Chen, J.~M. Yao, S.~Q. Zhang, B.~Qi,
\newblock Phys. Rev. C 82 (2010) 067302.

\bibitem{Zhang2007Chiral}
S.~Q. Zhang, B.~Qi, S.~Y. Wang, J.~Meng,
\newblock Phys. Rev. C 75 (2007) 044307.

\bibitem{Wang2008Description}
S.~Y. Wang, S.~Q. Zhang, B.~Qi, J.~Peng, J.~M. Yao, J.~Meng,
\newblock Phys. Rev. C 77 (2008) 034314.

\bibitem{Wang2010Theoretical}
S.~Y. Wang, B.~Qi, D.~P. Sun,
\newblock Phys. Rev. C 82 (2010) 027303.

\bibitem{Qi2011Chiral}
B.~Qi, S.~Y. Wang, S.~Q. Zhang,
\newblock Chin. Phys. Lett. 28 (2011) 122101.

\bibitem{Qi2009Chirality}
B.~Qi, S.~Q. Zhang, J.~Meng, S.~Y. Wang, S.~Frauendorf,
\newblock Phys. Lett. B 675 (2009) 175.

\bibitem{Qi2011Chirality}
B.~Qi, S.~Q. Zhang, S.~Y. Wang, J.~Meng, T.~Koike,
\newblock Phys. Rev. C 83 (2011) 034303.

\bibitem{Qi2012Transition}
B.~Qi, J.~Li, S.~Y. Wang, J.~Zhang, S.~Q. Zhang,
\newblock Chin. Phys. Lett. 29 (2012) 072101.

\bibitem{Starosta2002Role}
K.~Starosta, C.~J. Chiara, D.~B. Fossan, T.~Koike, T.~T.~S. Kuo,
  D.~R. LaFosse, S.~G. Rohozi\ifmmode~\acute{n}\else \'{n}\fi{}ski,
  Ch. Droste, T.~Morek, J.~Srebrny,
\newblock Phys. Rev. C 65 (2002) 044328.

\bibitem{Koike2003Systematic}
T.~Koike, K.~Starosta, C.~J. Chiara, D.~B. Fossan, D.~R. LaFosse,
\newblock Phys. Rev. C 67 (2003) 044319.

\bibitem{Droste2009Chiral}
Ch. Droste, S.~G. Rohozi$\mathrm{\acute{n}}$ski, K.~Starosta,
  L.~Pr$\mathrm{\acute{o}}$chniak, E.~Grodner,
\newblock Eur. Phys. J. A 42 (2009) 79.

\bibitem{Tonev2006}
D.~Tonev, G.~de~Angelis, P.~Petkov, A.~Dewald, S.~Brant,
  S.~Frauendorf, D.~L. Balabanski, P.~Pejovic, D.~Bazzacco,
  P.~Bednarczyk, F.~Camera, A.~Fitzler, A.~Gadea, S.~Lenzi,
  S.~Lunardi, N.~Marginean, O.~M\"oller, D.~R. Napoli, A.~Paleni,
  C.~M. Petrache, G.~Prete, K.~O. Zell, Y.~H. Zhang, Jing-ye Zhang,
  Q.~Zhong, D.~Curien,
\newblock Phys. Rev. Lett. 96 (2006) 052501.

\bibitem{Tonev2007}
D.~Tonev, G.~de Angelis, S.~Brant, S.~Frauendorf, P.~Petkov,
  A.~Dewald, F.~D\"onau, D.~L. Balabanski, Q.~Zhong, P.~Pejovic,
  D.~Bazzacco, P.~Bednarczyk, F.~Camera, D.~Curien, F.~Della Vedova,
  A.~Fitzler, A.~Gadea, G.~Lo Bianco, S.~Lenzi, S.~Lunardi,
  N.~Marginean, O.~M\"oller, D.~R. Napoli, R.~Orlandi, E.~Sahin,
  A.~Saltarelli, J.~Valiente Dobon, K.~O. Zell, Jing-ye Zhang,
  Y.~H. Zhang,
\newblock Phys. Rev. C 76 (2007) 044313.

\bibitem{Brant2008Dynamic}
S.~Brant, D.~Tonev, G.~De Angelis, A.~Ventura,
\newblock Phys. Rev. C 78 (2008) 034301.

\bibitem{Brant2009Chiral}
S.~Brant, C.~M. Petrache,
\newblock Phys. Rev. C 79 (2009) 054326.

\bibitem{Raduta2016New}
A.~A. Raduta, Al.~H. Raduta, C.~M. Petrache,
\newblock J. Phys. G: Nucl. Part. Phys. 43 (2016) 095107.

\bibitem{Shimada2018Rotational}
M.~Shimada, Y.~Fujioka, S.~Tagami, Y.~R. Shimizu,
\newblock Phys. Rev. C 97 (2018) 024319.

\bibitem{Higashiyama2005New}
K.~Higashiyama, N.~Yoshinaga, K.~Tanabe,
\newblock Phys. Rev. C 72 (2005) 024315.

\bibitem{Higashiyama2013Pair}
K.~Higashiyama, N.~Yoshinaga,
\newblock Phys. Rev. C 88 (2013) 034315.

\bibitem{Dimitrov2000Chirality}
V.~I. Dimitrov, S.~Frauendorf, F.~D$\mathrm{\ddot{o}}$nau,
\newblock Phys. Rev. Lett. 84 (2000) 5732.

\bibitem{Madokoro2000Relativistic}
H.~Madokoro, J.~Meng, M.~Matsuzaki, S.~Yamaji,
\newblock Phys. Rev. C 62 (2000) 061301(R).

\bibitem{Meng2013Progress}
J.~Meng, J.~Peng, S.~Q. Zhang, P.~W. Zhao,
\newblock Front. Phys. 8 (2013) 55.

\bibitem{Olbratowski2004Critical}
P.~Olbratowski, J.~Dobaczewski, J.~Dudek, W.~P\l{}\'ociennik,
\newblock Phys. Rev. Lett. 93 (2004) 052501.

\bibitem{Mukhopadhyay2007From}
S.~Mukhopadhyay, D.~Almehed, U.~Garg, S.~Frauendorf, T.~Li,
  P.~V.~Madhusudhana Rao, X.~Wang, S.~S. Ghugre,
  M.~P. Carpenter, S.~Gros, A.~Hecht, R.~V.~F.~Janssens,
  F.~G.~Kondev, T.~Lauritsen,~D.~Seweryniak, S. Zhu,
\newblock Phys. Rev. Lett. 99 (2007) 172501.

\bibitem{Almehed2011Chiral}
D.~Almehed, F.~D$\mathrm{\ddot{o}}$nau, S.~Frauendorf,
\newblock Phys. Rev. C 83 (2011) 054308.

\bibitem{Chen2013Collective}
Q.~B. Chen, S.~Q. Zhang, P.~W. Zhao, R.~V. Jolos, J.~Meng,
\newblock Phys. Rev. C 87 (2013) 024314.

\bibitem{Chen2016Two}
Q.~B. Chen, S.~Q. Zhang, P.~W. Zhao, R.~V. Jolos, J.~Meng,
\newblock Phys. Rev. C 94 (2016) 044301.

\bibitem{Hara1995PROJECTED}
K.~Hara, Y.~Sun,
\newblock Int. J. Mod. Phys. E 4 (1995) 637.

\bibitem{Bhat2014Investigation}
G.~H. Bhat, R.~N. Ali, J.~A. Sheikh, R.~Palit,
\newblock Nuclear Phys. A 922 (2014) 150.

\bibitem{Chen2017Chiral}
F.~Q. Chen, Q.~B. Chen, Y.~A. Luo, J.~Meng, S.~Q. Zhang,
\newblock Phys. Rev. C 96 (2017) 051303(R).

\bibitem{Chen2018APPB}
F.~Q. Chen, J.~Meng,
\newblock Acta Phys. Pol. B 11 (2018) 1001.

\bibitem{RING1996PPNP}
P.~Ring,
\newblock Prog. Part. Nucl. Phys. 37 (1996) 193.

\bibitem{MENG2006PPNP}
J.~Meng, H.~Toki, S.~G. Zhou, S.~Q. Zhang, W.~H. Long,
  L.~S. Geng,
\newblock Prog. Part. Nucl. Phys. 57 (2006) 470.

\bibitem{Meng2015JPG}
J.~Meng, S.~G. Zhou,
\newblock J. Phys. G: Nucl. Part. Phys. 42 (2015) 093101.

\bibitem{LIANG2015Hidden}
H.~Z. Liang, J.~Meng, S.~G. Zhou,
\newblock Phys. Rep. 570 (2015) 1.

\bibitem{Meng2006Possible}
J.~Meng, J.~Peng, S.~Q. Zhang, S.~G. Zhou,
\newblock Phys. Rev. C 73 (2006) 037303.

\bibitem{Yao2009Candidate}
J.~M. Yao, B.~Qi, S.~Q. Zhang, J.~Peng, S.~Y. Wang, J.~Meng,
\newblock Phys. Rev. C 79 (2009) 067302.

\bibitem{Peng2008Search}
J.~Peng, H.~Sagawa, S.~Q. Zhang, J.~M. Yao, Y.~Zhang, J.~Meng,
\newblock Phys. Rev. C 77 (2008) 024309.

\bibitem{Ayangeakaa2013Evidence}
A.~D. Ayangeakaa, U.~Garg, M.~D. Anthony, S.~Frauendorf,
  J.~T. Matta, B.~K. Nayak, D.~Patel, Q.~B. Chen, S.~Q. Zhang,
  P.~W. Zhao, B.~Qi, J.~Meng, R.~V.~F. Janssens,
  M.~P. Carpenter, C.~J. Chiara, F.~G. Kondev, T.~Lauritsen,
  D.~Seweryniak, S.~Zhu, S.~S. Ghugre, R.~Palit,
\newblock Phys. Rev. Lett. 110 (2013) 172504.

\bibitem{Kuti2014Multiple}
I.~Kuti, Q.~B. Chen, J.~Tim\'ar, D.~Sohler, S.~Q. Zhang,
  Z.~H. Zhang, P.~W. Zhao, J.~Meng, K.~Starosta, T.~Koike,
  E.~S. Paul, D.~B. Fossan, C.~Vaman,
\newblock Phys. Rev. Lett. 113 (2014) 032501.

\bibitem{Liu2016Evidence}
C.~Liu, S.~Y. Wang, R.~A. Bark, S.~Q. Zhang, J.~Meng, B.~Qi,
  P.~Jones, S.~M. Wyngaardt, J.~Zhao, C.~Xu, S.-G. Zhou,
  S.~Wang, D.~P. Sun, L.~Liu, Z.~Q. Li, N.~B. Zhang, H.~Jia,
  X.~Q. Li, H.~Hua, Q.~B. Chen, Z.~G. Xiao, H.~J. Li, L.~H. Zhu,
  T.~D. Bucher, T.~Dinoko, J.~Easton, K.~Juh\'asz, A.~Kamblawe,
  E.~Khaleel, N.~Khumalo, E.~A. Lawrie, J.~J. Lawrie,
  S.~N.~T. Majola, S.~M. Mullins, S.~Murray, J.~Ndayishimye,
  D.~Negi, S.~P. Noncolela, S.~S. Ntshangase, B.~M. Nyak\'o,
  J.~N. Orce, P.~Papka, J.~F. Sharpey-Schafer, O.~Shirinda,
  P.~Sithole, M.~A. Stankiewicz, M.~Wiedeking,
\newblock Phys. Rev. Lett. 116 (2016) 112501.

\bibitem{Petrache2018Evidence}
  C.~M. Petrache, B.~F. Lv, A.~Astier, E.~Dupont, Y.~K. Wang,
  S.~Q. Zhang, P.~W. Zhao, Z.~X. Ren, J.~Meng, P.~T. Greenlees,
  H.~Badran, D.~M. Cox, T.~Grahn, R.~Julin, S.~Juutinen, J.~Konki,
  J.~Pakarinen, P.~Papadakis, J.~Partanen, P.~Rahkila,
  M.~Sandzelius, J.~Saren, C.~Scholey, J.~Sorri, S.~Stolze,
  J.~Uusitalo, B.~Cederwall, \"O. Aktas, A.~Ertoprak, H.~Liu,
  S.~Matta, P.~Subramaniam, S.~Guo, M.~L. Liu, X.~H. Zhou,
  K.~L. Wang, I.~Kuti, J.~Tim\'ar, A.~Tucholski, J.~Srebrny,
  C.~Andreoiu,
\newblock Phys. Rev. C 97 (2018) 041304(R).

\bibitem{Alc2004Magnetic}
J.~A. Alc\'antara-N\'u\~nez, J.~R.~B. Oliveira, E.~W. Cybulska,
  N.~H. Medina, M.~N. Rao, R.~V. Ribas, M.~A. Rizzutto, W.~A. Seale,
  F.~Falla-Sotelo, K.~T. Wiedemann, V.~I. Dimitrov, S.~Frauendorf,
\newblock Phys. Rev. C 69 (2004) 024317.

\bibitem{Tim2004Experimental}
J.~Tim$\mathrm{\acute{a}}$r, P.~Joshi, K.~Starosta, V.~I. Dimitrov,
  D.~B. Fossan, J.~Moln$\mathrm{\acute{a}}$r, D.~Sohler,
  R.~Wadsworth, A.~Algora, P.~Bednarczyk, D.~Curieng,
  Zs. Dombr$\mathrm{\acute{a}}$di, G.~Duchene, A.~Gizon, J.~Gizon,
  D.~G. Jenkins, T.~Koike, A.~Krasznahorkay, E.~S. Paul, P.~M.
  Raddon, G.~Rainovski, J.~N. Scheurer, A.~J. Simons, C.~Vaman, A.~R.
  Wilkinson, L.~Zolnai, S.~Frauendorf,
\newblock Phys. Lett. B 598 (2004) 178.

\bibitem{Li2011Multiple}
J.~Li, S.~Q.~Zhang, J.~Meng,
\newblock Phys. Rev. C 83 (2011) 037301.

\bibitem{He2012Quest}
C.~Y. He, B.~Zhang, L.~H. Zhu, X.~G. Wu, H.~B. Sun, Y.~Zheng,
  B.~B. Yu, L.~L. Wang, G.~S. Li, S.~H. Yao, C.~Xu, J. G.~Wang,
  L.~Gu,
\newblock Plasma Sci. Technol. 14 (2012) 518.

\bibitem{Dan1994Collective}
Dan Jerrestam, W.~Klamra, J.~Gizon, F.~Lid$\mathrm{\acute{e}}$n,
  L.~Hildingsson, J.~Kownacki, Th. Lindblad, J.~Nyberg,
\newblock Nuclear Phys. A 577 (1994) 786.

\bibitem{Qi2013Possible}
B.~Qi, H.~Jia, N.~B. Zhang, C.~Liu, S.~Y. Wang,
\newblock Phys. Rev. C 88 (2013) 027302.

\bibitem{Petrache2012Tilted}
C.~M. Petrache, S.~Frauendorf, M.~Matsuzaki, R.~Leguillon,
  T.~Zerrouki, S.~Lunardi, D.~Bazzacco, C.~A. Ur, E.~Farnea,
  C.~Rossi~Alvarez, R.~Venturelli, G.~de~Angelis,
\newblock Phys. Rev. C 86 (2012) 044321.

\bibitem{Peng2008PRC}
J.~Peng, J.~Meng, P.~Ring, S.~Q. Zhang,
\newblock Phys. Rev. C 78 (2008) 024313.

\bibitem{ZHAO2011PLB}
P.~W. Zhao, S.~Q. Zhang, J.~Peng, H.~Z. Liang, P.~Ring, J.~Meng,
\newblock Phys. Lett. B 699 (2011) 181.

\bibitem{Yu2012PRC}
L.~F. Yu, P.~W. Zhao, S.~Q. Zhang, P.~Ring, J.~Meng,
\newblock Phys. Rev. C 85 (2012) 024318.

\bibitem{Zhao2011PRL}
P.~W. Zhao, J.~Peng, H.~Z. Liang, P.~Ring, J.~Meng,
\newblock Phys. Rev. Lett. 107 (2011) 122501.

\bibitem{Zhao2012PRC}
P.~W. Zhao, J.~Peng, H.~Z. Liang, P.~Ring, J.~Meng,
\newblock Phys. Rev. C 85 (2012) 054310.

\bibitem{Zhao2015Impact}
P.~W. Zhao, S.~Q. Zhang, J.~Meng,
\newblock Phys. Rev. C 92 (2015) 034319.

\bibitem{Wang2017Yrast}
Y.~K. Wang,
\newblock Phys. Rev. C 96 (2017) 054324.

\bibitem{Zhao2015PRL}
P.~W. Zhao, N.~Itagaki, J.~Meng,
\newblock Phys. Rev. Lett. 115 (2015) 022501.

\bibitem{Zhao2017Multiple}
P.~W. Zhao,
\newblock Phys. Lett. B 773 (2017) 1.

\bibitem{Wang2007Examining}
S.~Y. Wang, S.~Q. Zhang, B.~Qi, J.~Meng,
\newblock Chin. Phys. Lett. 24 (2007) 664.

\bibitem{Wang2011The}
S.~Y. Wang, B.~Qi, L.~Liu, S.~Q. Zhang, H.~Hua, X.~Q. Li,
  Y.~Y. Chen, L.~H. Zhu, J.~Meng, S.~M. Wyngaardt, P.~Papka,
  T.~T. Ibrahim, R.~A. Bark, P.~Datta, E.~A. Lawrie,
  J.~J. Lawrie, S.~N.~T. Majola, P.~L. Masiteng, S.~M. Mullins,
  J.~G$\mathrm{\acute{a}}$l, G.~Kalinka,
  J.~Moln$\mathrm{\acute{a}}$r, B.~M. Nyak$\mathrm{\acute{o}}$,
  J.~Tim$\mathrm{\acute{a}}$r, K.~Juh$\mathrm{\acute{a}}$sz,
  R.~Schwengner,
\newblock Phys. Lett. B 703 (2011) 40.

\bibitem{Zhu2009Search}
S.~J. Zhu, J.~H. Hamilton, A.~V. Ramayya, J.~K. Hwang, J.~O.
  Rasmussen, Y.~X. Luo, K.~Li, J.~G. Wang, X.~L. Che, H.~B. Ding,
  S.~Frauendorf, V.~Dimitrov, Q.~Xu, L.~Gu, Y.~Y. Yang,
\newblock Chin. Phys. C 33 (2009) 145.

\bibitem{Ding2010Proposed}
H.~B. Ding, S.~J. Zhu, J.~G. Wang, L.~Gu, Q.~Xu, Z.~G. Xiao,
  E. Y. Yeoha, M.~Zhang, L.~H. Zhu, X.~G. Wu, Y.~Liu, C.~Y. He,
  L.~L. Wang, B.~Pan, G.~S. Li,
\newblock Chin. Phys. Lett. 27 (2010) 072501.

\bibitem{Joshi2005First}
P.~Joshi, A.~R. Wilkinson, T.~Koike, D.~B. Fossan, S.~Finnigan,
  E.~S. Paul, P.~M. Raddon, G.~Rainovski, K.~Starosta,
  A.~J. Simons, C.~Vaman, R.~Wadsworth,
\newblock Eur. Phys. J. A 24 (2005) 23.

\bibitem{Y2009ODD}
Y.~X. Luo, S.~J. Zhu, J.~H. Hamilton, A.~V. Ramayya, C.~Goodin,
  K.~Li, X.~L. Che, J.~K. Hwang, I.~Y. Lee, Z.~Jiang, G.~M.
  Ter-akopian, A.~V. Daniel, M.~A. Stoyer, R.~Donangelo,
  S.~Frauendorf, V.~Dimitrov, Jing-ye Zhang, J.~D. Cole,
  N.~J. Stone, J.~O. Rasmussen,
\newblock Int. J. Mod. Phys. E 18 (2009) 1697.

\bibitem{Tonev2014Candidates}
D.~Tonev, M.~S. Yavahchova, N.~Goutev, G.~de~Angelis, P.~Petkov,
  R.~K. Bhowmik, R.~P. Singh, S.~Muralithar, N.~Madhavan,
  R.~Kumar, M.~Kumar~Raju, J.~Kaur, G.~Mohanto, A.~Singh,
  N.~Kaur, R.~Garg, A.~Shukla, Ts.~K. Marinov, S.~Brant.
\newblock Phys. Rev. Lett. 112 (2014) 052501.

\bibitem{Vaman2004Chiral}
C.~Vaman, D.~B. Fossan, T.~Koike, K.~Starosta, I.~Y. Lee, A.~O.
  Macchiavelli,
\newblock Phys. Rev. Lett. 92 (2004) 032501.

\bibitem{Joshi2004Stability}
P.~Joshi, D.~G. Jenkins, P.~M. Raddon, A.~J. Simons,
  R.~Wadsworth, A.~R. Wilkinson, D.~B. Fossan, T.~Koike,
  K.~Starosta, C.~Vaman, J.~Tim$\mathrm{\acute{a}}$r,
  Zs. Dombr$\mathrm{\acute{a}}$di, A.~Krasznahorkay,
  J.~Moln$\mathrm{\acute{a}}$r, D.~Sohler, L.~Zolnai, A.~Algora,
  E.~S. Paul, G.~Rainovski, A.~Gizon, J.~Gizon, P.~Bednarczyk,
  D.~Curien, G.~Duch$\mathrm{\hat{e}}$ne, J.~N. Scheurer,
\newblock Phys. Lett. B 595 (2004) 135.

\bibitem{Luo2004Level}
Y.~X. Luo, S.~C. Wu, J.~Gilat, J.~O. Rasmussen, J.~H. Hamilton,
  A.~V. Ramayya, J.~K. Hwang, C.~J. Beyer, S.~J. Zhu, J.~Kormicki,
  X.~Q. Zhang, E.~F. Jones, P.~M. Gore, I-Yang Lee, P.~Zielinski,
  C.~M. Folden, T.~N. Ginter, P.~Fallon, G.~M. Ter-Akopian,
  A.~V. Daniel, M.~A. Stoyer, J.~D. Cole, R.~Donangelo, S.~J.
  Asztalos, A.~Gelberg,
\newblock Phys. Rev. C 69 (2004) 024315.

\bibitem{Wang2013High}
Z.~G. Wang, M.~L. Liu, Y.~H. Zhang, X.~H. Zhou, B.~T. Hu, N.~T.
  Zhang, S.~Guo, B.~Ding, Y.~D. Fang, J.~G. Wang, G.~S. Li, Y.~H.
  Qiang, S.~C. Li, B.~S. Gao, Y.~Zheng, W.~Hua, X.~G. Wu, C.~Y. He,
  Y.~Zheng, C.~B. Li, J.~J. Liu, S.~P. Hu.
\newblock Phys. Rev. C 88 (2013) 024306.

\bibitem{Tim2007High}
J.~Tim$\mathrm{\acute{a}}$r, T.~Koike, N.~Pietralla, G.~Rainovski,
  D.~Sohler, T.~Ahn, G.~Berek, A.~Costin, K.~Dusling, T. C.~Li,
  E. S.~Paul, K.~Starosta, C.~Vaman,
\newblock Phys. Rev. C 76 (2007) 024307.

\bibitem{Lieder2014Resolution}
E.~O. Lieder, R.~M. Lieder, R.~A. Bark, Q.~B. Chen, S.~Q. Zhang,
  J.~Meng, E.~A. Lawrie, J.~J. Lawrie, S.~P. Bvumbi, N.~Y. Kheswa,
  S.~S. Ntshangase, T.~E. Madiba, P.~L. Masiteng, S.~M. Mullins,
  S.~Murray, P.~Papka, D.~G. Roux, O.~Shirinda, Z.~H. Zhang,
  P.~W. Zhao, Z.~P. Li, J.~Peng, B.~Qi, S.~Y. Wang, Z.~G. Xiao,
  C.~Xu,
\newblock Phys. Rev. Lett. 112 (2014) 202502.

\bibitem{Starosta2001Chirality}
K.~Starosta, T.~Koike, C.~J. Chiara, D.~B. Fossan, D.~R. LaFosse,
\newblock Nuclear Phys. A 682 (2001) 375c.

\bibitem{Zhao2009Observation}
Y.~X. Zhao, T.~Komatsubara, Y.~J. Ma, Y.~H. Zhang, S~.Y. Wang,
  Y.~Z. Liu, K.~Furuno,
\newblock Chin. Phys. Lett. 26 (2009) 082301.

\bibitem{Zheng2012Abnormal}
Y.~Zheng, L.~H. Zhu, X.~G. Wu, Z.~C. Gao, C.~Y. He, G.~S. Li,
  L.~L. Wang, Y.~S. Chen, Y.~Sun, X.~Hao, Y.~Liu, X.~Q. Li,
  B.~Pan, Y.~J. Ma, Z.~Y. Li, H.~B. Ding,
\newblock Phys. Rev. C 86 (2012) 014320.

\bibitem{Yonnam2005}
U.~Yon-Nam, S.~J. Zhu, M.~Sakhaee, L.~M. Yang, C.~Y. Gan, L.~Y.
  Zhu, R.~Q. Xu, X.~L. Che, M.~L. Li, Y.~J. Chen, S.~X. Wen,
  X.~G. Wu, L.~H. Zhu, G.~S. Li, J.~Peng, S.~Q. Zhang, J.~Meng,
\newblock J. Phys. G: Nucl. Part. Phys. 31 (2005) B1.

\bibitem{Selvakumar2015Evidence}
K.~Selvakumar, A.~K. Singh, Chandan Ghosh, Purnima Singh,
  A.~Goswami, R.~Raut, A.~Mukherjee, U.~Datta, P.~Datta, S.~Roy,
  G.~Gangopadhyay, S.~Bhowal, S.~Muralithar, R.~Kumar, R.~P.
  Singh, M.~Kumar Raju,
\newblock Phys. Rev. C 92 (2015) 064307.

\bibitem{Grodner2011Partner}
E.~Grodner, I.~Sankowska, T.~Morek, S.~G.
  Rohozi$\mathrm{\acute{n}}$ski, Ch. Droste, J.~Srebrny,
  A.~A. Pasternak, M.~Kisieli$\mathrm{\acute{n}}$ski,
  M.~Kowalczyk, J.~Kownacki, J.~Mierzejewski,
  A.~Kr$\mathrm{\acute{o}}$l, K.~Wrzosek,
\newblock Phys. Lett. B 703 (2011) 46.

\bibitem{Grodner2006128Cs}
E.~Grodner, J.~Srebrny, A.~A. Pasternak, I.~Zalewska, T.~Morek,
  Ch. Droste, J.~Mierzejewski, M.~Kowalczyk, J.~Kownacki,
  M.~Kisieli\ifmmode~\acute{n}\else\'{n}\fi{}ski, S.~G.
  Rohozi\ifmmode~\acute{n}\else \'{n}\fi{}ski, T.~Koike,
  K.~Starosta, A.~Kordyasz, P.~J. Napiorkowski, M.~Woli\'{n}ska-
  Cichocka, E.~Ruchowska, W.~P\l{}\'ociennik, J.~Perkowski,
\newblock Phys. Rev. Lett. 97 (2006) 172501.

\bibitem{Simons2005Evidence}
A.~J. Simons, P.~Joshi, D.~G. Jenkins, P.~M. Raddon, R.~Wadsworth,
  D.~B. Fossan, T.~Koike, C.~Vaman, K.~Starosta, E.~S. Paul,
  H.~J. Chantler, A.~O. Evans, P.~Bednarczyk, D.~Curien,
\newblock J. Phys. G: Nucl. Part. Phys. 31 (2005) 541.

\bibitem{Rainovski2003Candidate}
G.~Rainovski, E.~S. Paul, H.~J. Chantler, P.~J. Nolan, D.~G.
  Jenkins, R.~Wadsworth, P.~Raddon, A.~Simons, D.~B. Fossan,
  T.~Koike, K.~Starosta, C.~Vaman, E.~Farnea, A.~Gadea,
  Th. Kr\"oll, R.~Isocrate, G.~de Angelis, D.~Curien,
  V.~I. Dimitrov,
\newblock Phys. Rev. C 68 (2003) 024318.

\bibitem{Ma2012Candidate}
K.~Y. Ma, J.~B. Lu, D.~Yang, H.~D. Wang, Y.~Z. Liu, X.~G. Wu,
  Y.~Zheng, C.~Y. He,
\newblock Phys. Rev. C 85 (2012) 037301.

\bibitem{Koike2001Observation}
T.~Koike, K.~Starosta, C.~J. Chiara, D.~B. Fossan, D.~R.
  LaFosse,
\newblock Phys. Rev. C 63 (2001) 061304(R).

\bibitem{Kuti2013Medium}
I.~Kuti, J.~Tim\'ar, D.~Sohler, E.~S. Paul, K.~Starosta,
  A.~Astier, D.~Bazzacco, P.~Bednarczyk, A.~J. Boston,
  N.~Buforn, H.~J. Chantler, C.~J. Chiara, R.~M. Clark,
  M.~Cromaz, M.~Descovich, Zs. Dombr\'adi, P.~Fallon,
  D.~B. Fossan, C.~Fox, A.~Gizon, J.~Gizon, A.~A. Hecht,
  N.~Kintz, T.~Koike, I.~Y. Lee, S.~Lunardi, A.~O.
  Macchiavelli, P.~J. Nolan, B.~M. Nyak\'o, C.~M. Petrache,
  J.~A. Sampson, H.~C. Scraggs, T.~G. Tornyi, R.~Wadsworth,
  A.~Walker, L.~Zolnai,
\newblock Phys. Rev. C 87 (2013) 044323.

\bibitem{Petrache2016Triaxial}
C.~M. Petrache, Q.~B. Chen, S.~Guo, A.~D. Ayangeakaa,
  U.~Garg, J.~T. Matta, B.~K. Nayak, D.~Patel, J.~Meng,
  M.~P. Carpenter, C.~J. Chiara, R.~V.~F. Janssens,
  F.~G. Kondev, T.~Lauritsen, D.~Seweryniak, S.~Zhu,
  S.~S. Ghugre, R.~Palit,
\newblock Phys. Rev. C 94 (2016) 064309.

\bibitem{Bark2001Candidate}
R.~A. Bark, A.~M. Baxter, A.~P. Byrne, G.~D. Dracoulis,
  T.~Kib$\mathrm{\acute{e}}$di, T.~R. McGoram, S.~M.
  Mullins,
\newblock Nuclear Phys. A 691 (2001) 577.

\bibitem{Timar2011Medium}
J.~Tim\'ar, K.~Starosta, I.~Kuti, D.~Sohler, D.~B. Fossan,
  T.~Koike, E.~S. Paul, A.~J. Boston, H.~J. Chantler,
  M.~Descovich, R.~M. Clark, M.~Cromaz, P.~Fallon, I.~Y.
  Lee, A.~O. Macchiavelli, C.~J. Chiara, R.~Wadsworth, A.~A.
  Hecht, D.~Almehed, S.~Frauendorf,
\newblock Phys. Rev. C 84 (2011) 044302.

\bibitem{Petrache1997High}
C.~M. Petrache, R.~Venturelli, D.~Vretenar, D.~Bazzacco,
  G.~Bonsignori, S.~Brant, S.~Lunardi, M.~A. Rizzutto,
  C.~Rossi Alvarez, G.~de Angelis, M.~De Poli, D.~R. Napoli,
\newblock Nuclear Phys. A 617 (1997) 228.

\bibitem{Hartley2001Detailed}
D.~J. Hartley, L.~L. Riedinger, M.~A. Riley, D.~L. Balabanski,
  F.~G. Kondev, R.~W. Laird, J.~Pfohl, D.~E. Archer, T.~B.
  Brown, R.~M. Clark, M.~Devlin, P.~Fallon, I.~M. Hibbert,
  D.~T. Joss, D.~R. LaFosse, P.~J. Nolan, N.~J. O'Brien, E.~S.
  Paul, D.~G. Sarantites, R.~K. Sheline, S.~L. Shepherd, J.~
  Simpson, R.~Wadsworth, Jing-ye Zhang, P.~B. Semmes,
  F.~D\"onau,
\newblock Phys. Rev. C 64 (2001) 031304(R).

\bibitem{Ma2018Candidate}
K.~Y. Ma, J.~B. Lu, Z.~Zhang, J.~Q. Liu, D.~Yang, Y.~M. Liu,
  X.~Xu, X.~Y. Li, Y.~Z. Liu, X.~G. Wu, Y.~Zheng, C.~B. Li,
\newblock Phys. Rev. C 97 (2018) 014305.

\bibitem{Hecht2001Evidence}
A.~A. Hecht, C.~W. Beausang, K.~E. Zyromski, D.~L. Balabanski,
  C.~J. Barton, M.~A. Caprio, R.~F. Casten, J.~R. Cooper, D.~J.
  Hartley, R.~Kr\"ucken, D.~Meyer, H.~Newman, J.~R. Novak, E.~S.
  Paul, N.~Pietralla, A.~Wolf, N.~V. Zamfir, Jing-ye Zhang,
  F.~D\"onau,
\newblock Phys. Rev. C 63 (2001) 051302(R).

\bibitem{Hecht2003Evidence}
A.~A. Hecht, C.~W. Beausang, H.~Amro, C.~J. Barton, Z.~Berant,
  M.~A. Caprio, R.~F. Casten, J.~R. Cooper, D.~J. Hartley,
  R.~Kr\"ucken, D.~A. Meyer, H.~Newman, J.~R. Novak,
  N.~Pietralla, J.~J. Ressler, A.~Wolf, N.~V. Zamfir, Jing-ye
  Zhang, K.~E. Zyromski,
\newblock Phys. Rev. C 68 (2003) 054310.

\bibitem{Balabanski2004Possible}
D.~L. Balabanski, M.~Danchev, D.~J. Hartley, L.~L. Riedinger,
  O.~Zeidan, Jing-ye Zhang, C.~J. Barton, C.~W. Beausang, M.~A.
  Caprio, R.~F. Casten, J.~R. Cooper, A.~A. Hecht, R.~Kr\"ucken,
  J.~R. Novak, N.~V. Zamfir, K.~E. Zyromski,
\newblock Phys. Rev. C 70 (2004) 044305.

\bibitem{Ndayishimye2017Chiral}
J.~Ndayishimye, E.~A. Lawrie, O.~Shirinda, J.~L. Easton, S.~M.
  Wyngaardt, R.~A. Bark, S.~P. Bvumbi, T.~R.~S. Dinoko, P.~Jones,
  N.~Y. Kheswa, J.~J. Lawrie, S.~N.~T. Majola, P.~L. Masiteng,
  D. Negi, J.~N.~Orce, P.~Papka, J.~F. Sharpey-Schafer,
  M.~Stankiewicz, M.~Wiedeking,
\newblock Acta Phys. Pol. B 48 (2017) 343.

\bibitem{Masiteng2014Rotational}
P.~L. Masiteng, E.~A. Lawrie, T.~M. Ramashidzha, J.~J. Lawrie,
  R.~A. Bark, R.~Lindsay, F.~Komati, J.~Kau, P.~Maine, S.~M.
  Maliage, I. Matamba, S.~M. Mullins, S.~H.~T. Murray, K.~P.
  Mutshena, A.~A. Pasternak, D.~G. Roux, J.~F. Sharpey-Schafer,
  O.~Shirinda, P.~A. Vymers,
\newblock Eur. Phys. J. A 50 (2014) 119.

\bibitem{Lawrie2010Candidate}
E.~A. Lawrie, P.~A. Vymers, Ch. Vieu, J.~J. Lawrie, C.~Sch\"uck,
  R.~A. Bark, R.~Lindsay, G.~K. Mabala, S.~M. Maliage, P.~L.
  Masiteng, S.~M. Mullins, S.~H.~T. Murray, I.~Ragnarsson, T.~M.
  Ramashidzha, J.~F. Sharpey-Schafer, O.~Shirinda,
\newblock Eur. Phys. J. A 45 (2010) 39.

\bibitem{Zhu2005Soft}
S.~J. Zhu, J.~H. Hamilton, A.~V. Ramayya, P.~M. Gore, J.~O.
  Rasmussen, V.~Dimitrov, S.~Frauendorf, R.~Q. Xu, J.~K. Hwang,
  D.~Fong, L.~M. Yang, K.~Li, Y.~J. Chen, X.~Q. Zhang, E.~F.
  Jones, Y.~X. Luo, I.~Y. Lee, W.~C. Ma, J.~D. Cole, M.~W.
  Drigert, M.~Stoyer, G.~M. Ter-Akopian, A.~V. Daniel,
\newblock Eur. Phys. J. A 25 (Suppl. 1) (2005) 459.

\bibitem{Luo2009Evolution}
Y.~X. Luo, S.~J. Zhu, J.~H. Hamilton, J.~O. Rasmussen, A.~V.
  Ramayya, C.~Goodin, K.~Li, J.~K. Hwang, D.~Almehed,
  S.~Frauendorf, V.~Dimitrov, Jing-ye Zhang, X.~L. Che, Z.~Jang,
  I.~Stefanescu, A.~Gelberg, G.~M. Ter-Akopian, A.~V. Daniel,
  M.~A. Stoyer, R.~Donangelo, J.~D. Cole, N.~J. Stone,
\newblock Phys. Lett. B 670 (2009) 307.

\bibitem{Frauendorf1996Interpretation}
S.~Frauendorf, J.~Meng,
\newblock Z. Phys. A 356 (1996) 263.

\bibitem{Zhang1999High}
Y.~H. Zhang, S.~Q. Zhang, Q.~Z. Zhao, S.~F. Zhu, H.~S. Xu, X.~H.
  Zhou, Y.~X. Guo, X.~G. Lei, J.~Lu, W.~X. Huang, Q.~B. Gou,
  H.~J. Jin, Z.~Liu, Y.~X. Luo, X.~F. Sun, Y.~T. Zhu, X.~G. Wu,
  S.~X. Wen, C.~X. Yang,
\newblock Phys. Rev. C 60 (1999) 044311.

\bibitem{Liu2018Br}
C.~Liu,
\newblock Private Communication, January 2018.

\bibitem{Hamilton2009New}
J.~H. Hamilton, Y.~X. Luo, S.~J. Zhu, J.~O. Rasmussen, A.~V.
  Ramayya, C.~Goodin, K.~Li, J.~K. Hwang, S.~Liu, D.~Almehed,
  S.~Frauendorf, V.~Dimitrov, Jing-ye Zhang, X.~L. Che, Z.~Jang,
  I.~Stefanescu, A.~Gelberg, G.~M. Ter-Akopian, A.~V. Daniel,
  I.~Y. Lee, H.~B. Ding, R.~Q. Xu, J.~G. Wang, Q.~Xu, M.~A.
  Stoyer, R.~Donangelo, N.~J. Stone,
\newblock Acta Phys. Pol. B 40 (2009) 523.

\bibitem{Vaman2004PhD}
C.~Vaman,
\newblock PhD Dissertation, State University of New York, 2004.

\bibitem{Timar2006Role}
J.~Tim\'ar, C.~Vaman, K.~Starosta, D.~B. Fossan, T.~Koike,
  D.~Sohler, I.~Y. Lee, A.~O. Macchiavelli,
\newblock Phys. Rev. C 73 (2006) 011301(R).

\bibitem{Suzuki2008Lifetime}
T.~Suzuki, G.~Rainovski, T.~Koike, T.~Ahn, M.~P. Carpenter,
  A.~Costin, M.~Danchev, A.~Dewald, R.~V.~F. Janssens, P.~Joshi,
  C.~J.~Lister, O.~M\"oller, N.~Pietralla, T.~Shinozuka,
  J.~Tim\'ar, R.~Wadsworth, C.~Vaman, S. Zhu,
\newblock Phys. Rev. C 78 (2008) 031302(R).

\bibitem{Timar2018Rh}
J.~Tim$\mathrm{\acute{a}}$r,
\newblock Private Communication, December 2017.

\bibitem{Joshi2005Evidence}
P.~Joshi, S.~Finnigan, D.~B. Fossan, T.~Koike, E.~S. Paul,
  G.~Rainovski, K.~Starosta, C.~Vaman, R.~Wadsworth,
\newblock J. Phys. G: Nucl. Part. Phys. 31 (2005) S1895.

\bibitem{Joshi2007Effect}
P.~Joshi, M.~P. Carpenter, D.~B. Fossan, T.~Koike, E.~S. Paul,
  G.~Rainovski, K.~Starosta, C.~Vaman, R.~Wadsworth,
\newblock Phys. Rev. Lett. 98 (2007) 102501.

\bibitem{Zheng2014Electromagnetic}
Y.~Zheng, L.~H. Zhu, X.~G. Wu, C.~Y. He, G.~S. Li, X.~Hao,
  B.~B. Yu, S.~H. Yao, B.~Zhang, C.~Xu, J.~G. Wang, L.~Gu,
\newblock Chin. Phys. Lett. 31 (2014) 062101.

\bibitem{Rather2014Exploring}
N.~Rather, P.~Datta, S.~Chattopadhyay, S.~Rajbanshi, A.~Goswami,
  G.~H. Bhat, J.~A. Sheikh, S.~Roy, R.~Palit, S.~Pal, S.~Saha,
  J.~Sethi, S.~Biswas, P.~Singh, H.~C. Jain,
\newblock Phys. Rev. Lett. 112 (2014) 202503.

\bibitem{R2014Studies}
R.~A. Bark, E.~O. Lieder, R.~M. Lieder, E.~A. Lawrie, J.~J.
  Lawrie, S.~P. Bvumbi, N.~Y. Kheswa, S.~S. Ntshangase, T.~E.
  Madiba, P.~L.~Masiteng, S. M.~Mullins, S. Murray, P.~Papka,
  O.~Shirinda, Q.~B. Chen, S.~Q. Zhang, Z.~H. Zhang, P.~W. Zhao,
  C.~Xu, J.~Meng, D.~G. Roux, Z.~P. Li, J.~Peng, B.~Qi,
  S.~Y. Wang, Z.~G. Xiao,
\newblock Int. J. Mod. Phys. E 23 (2014) 1461001.

\bibitem{Zhang2011New}
B.~Zhang, L.~H. Zhu, H.~B. Sun, C.~Y. He, X.~G. Wu, J.~B. Lu,
  Y.~J. Ma, X.~Hao, Y.~Zheng, B.~B. Yu, G.~S. Li, S.~H. Yao,
  L.~L. Wang, C.~Xu, J.~G. Wang, L.~Gu,
\newblock Chin. Phys. C 35 (2011) 1009.

\bibitem{Yao2014Lifetime}
S.~H. Yao, H.~L. Ma, L.~H. Zhu, X.~G. Wu, C.~Y. He, Y.~Zheng,
  B.~Zhang, G.~S. Li, C.~B. Li, S.~P. Hu, X.~P. Cao, B.~B. Yu,
  C.~Xu, Y.~Y. Cheng,
\newblock Phys. Rev. C 89 (2014) 014327.

\bibitem{Li2002Search}
X.~F. Li, Y.~J. Ma, Y.~Z. Liu, J.~B. Lu, G.~Y. Zhao, L.~C. Yin,
  R.~Meng, Z.~L. Zhang, L.~J. Wen, X.~H. Zhou, Y.~X. Guo, X.~G.
  Lei, Z.~Liu, J.~J. He, Y.~Zheng,
\newblock Chin. Phys. Lett. 19 (2002) 1779.

\bibitem{Wang2006Candidate}
S.~Y. Wang, Y.~Z. Liu, T.~Komatsubara, Y.~J. Ma, Y.~H. Zhang,
\newblock Phys. Rev. C 74 (2006) 017302.

\bibitem{Starosta2001Chiral}
K.~Starosta, T.~Koike, C.~J. Chiara, D.~B. Fossan, D.~R.
  LaFosse, A.~A. Hecht, C.~W. Beausang, M.~A. Caprio, J.~R.
  Cooper, R.~Kr\"ucken, J.~R. Novak, N.~V. Zamfir, K.~E.
  Zyromski, D.~J. Hartley, D.~L. Balabanski, Jing-ye Zhang,
  S.~Frauendorf, V.~I. Dimitrov,
\newblock Phys. Rev. Lett. 86 (2001) 971.

\bibitem{wang2009Lifetime}
L.~L. Wang, X.~G. Wu, L.~H. Zhu, G.~S. Li, X.~Hao, Y.~Zheng,
  C.~Y. He, L.~Wang, X.~Q. Li, Y.~Liu, B.~Pan, Z.~Y.~Li,
  H.~B. Ding,
\newblock Chin. Phys. C 33 (2009) 173.

\bibitem{Wu2012Test}
X.~G. Wu, L.~L. Wang, L.~H. Zhu, G.~S. Li, X.~Hao, Y.~Zheng,
  C.~Y. He, X.~Q. Li, B.~Pan, Y.~Liu, L.~Wang, Y.~X. Zhao,
  Z.~Y. Li, H.~B. Ding,
\newblock Plasma Sci. Technol. 14 (2012) 526.

\bibitem{Rainovski2003Planar}
G.~Rainovski, E.~S. Paul, H.~J. Chantler, P.~J. Nolan, D.~G.
  Jenkins, R.~Wadsworth, P.~Raddon, A.~Simons, D.~B. Fossan,
  T.~Koike, K.~Starosta, C.~Vaman, E.~Farnea, A.~Gadea, Th.
  Kr\"oll, G.~de Angelis, R.~Isocrate, D.~Curien, V.~I.
  Dimitrov,
\newblock J. Phys. G: Nucl. Part. Phys. 29 (2003) 2763.

\bibitem{Petrache1998Detailed}
C.~M. Petrache, S.~Brant, D.~Bazzacco, G.~Falconi, E.~Farnea,
  S.~Lunardi, V.~Paar, Zs. Podoly$\mathrm{\acute{a}}$k,
  R.~Venturelli, D.~Vretenar,
\newblock Nuclear Phys. A 635 (1998) 361.

\bibitem{Petrache1996Rotational}
C.~M. Petrache, D.~Bazzacco, S.~Lunardi, C.~Rossi Alvarez,
  G.~de Angelis, M.~De Poli, D.~Bucurescu, C.~A. Ur, P.~B.
  Semmes, R.~Wyss,
\newblock Nuclear Phys. A, 597 (1996) 106.

\bibitem{Petrache2006Risk}
C.~M. Petrache, G.~B. Hagemann, I.~Hamamoto, K.~Starosta,
\newblock Phys. Rev. Lett. 96 (2006) 112502.

\bibitem{D2006Lifetime}
D.~Tonev, P.~Petkov, D.~L. Balabanski, G.~de Angelis,
  A.~Gadea, D.~R. Napoli, N.~Marginean, A.~Dewald, P.~Pejovic,
  A.~Fitzler, O.~M$\mathrm{\ddot{o}}$ller, K.~O. Zell,
  S.~Brant, S.~Frauendorf, D.~Bazzacco, S.~Lenzi, S.~Lunardi,
  P.~Bednarczyk, D.~Curien, C.~Petrache, Q.~Zhong, Y.~H. Zhang,
  Jing-ye Zhang,
\newblock Int. J. Mod. Phys. E 15 (2006) 1531.

\bibitem{Tonev2007Question}
D.~Tonev, G.~de Angelis, S.~Brant, S.~Frauendorf, P.~Petkov,
  A.~Dewald, F.~D\"onau, D.~L. Balabanski, Q.~Zhong, P.~Pejovic,
  D.~Bazzacco, P.~Bednarczyk, F.~Camera, D.~Curien, F.~Della
  Vedova, A.~Fitzler, A.~Gadea, G.~Lo Bianco, S.~Lenzi, S.~
  Lunardi, N.~Marginean, O.~M\"oller, D.~R. Napoli, R.~Orlandi,
  E.~Sahin, A.~Saltarelli, J.~Valiente Dobon, K.~O. Zell, Jing-ye
  Zhang, Y.~H. Zhang,
\newblock Phys. Rev. C 76 (2007) 044313.

\bibitem{Zhu2003A}
S.~Zhu, U.~Garg, B.~K. Nayak, S.~S. Ghugre, N.~S. Pattabiraman,
  D.~B. Fossan, T.~Koike, K.~Starosta, C.~Vaman, R.~V.~F. Janssens,
  R.~S. Chakrawarthy, M.~Whitehead, A.~O. Macchiavelli, S.~Frauendorf,
\newblock Phys. Rev. Lett. 91 (2003) 132501.

\bibitem{Mergel2002Candidates}
E.~Mergel, C.~M. Petrache, G.~Lo~Bianco, H.~H{\"u}bel, J.~Domscheit,
  D.~Ro{\ss}bach, G.~Sch{\"o}nwa{\ss}er, N.~Nenoff, A.~Neu{\ss}er,
  A.~G{\"o}rgen, F.~Becker, E.~Bouchez, M.~Houry, A.~H{\"u}rstel,
  Y.~Le~Coz, R.~Lucas, Ch. Theisen, W.~Korten, A.~Bracco, N.~Blasi,
  F.~Camera, S.~Leoni, F.~Hannachi, A.~Lopez-Martens, M.~Rejmund,
  D.~Gassmann, P.~Reiter, P.~G. Thirolf, A.~Astier, N.~Buforn,
  M.~Meyer, N.~Redon, O.~Stezowski,
\newblock Eur. Phys. J. A 15 (2002) 417.

\bibitem{Mukhopadhyay2008Electromagnetic}
S.~Mukhopadhyay, D.~Almehed, U.~Garg, S.~Frauendorf, T.~Li, P.~V.
  Madhusudhana~Rao, X.~Wang, S.~S. Ghugre, M.~P. Carpenter, S.~Gros,
  A.~Hecht, R.~V.~F. Janssens, F.~G. Kondev, T.~Lauritsen,
  D.~Seweryniak, S.~Zhu,
\newblock Phys. Rev. C 78 (2008) 034311.

\bibitem{Beausang2001Evidence}
C.~W. Beausang, A.~A. Hecht, K.~E. Zyromski, D.~Balabanski, C.~J.~
  Barton, M.~A.~Caprio, R.~F. Casten, J.~R. Cooper, D.~Hartley,
  R.~Kr$\mathrm{\ddot{u}}$cken, J.~R. Novak, N.~V.~Zamfir,
  Jing-ye~Zhang, F.~D$\mathrm{\ddot{o}}$nau,
\newblock Nuclear Phys. A 682 (2001) 394c.

\bibitem{Hecht2004PhD}
A.~A. Hecht,
\newblock PhD Dissertation, Yale University, 2004.

\bibitem{Masiteng2013Close}
P.~L. Masiteng, E.~A. Lawrie, T.~M. Ramashidzha, R.~A. Bark, B.~G.
  Carlsson, J.~J. Lawrie, R.~Lindsay, F.~Komati, J.~Kau, P.~Maine,
  S.~M. Maliage, I.~Matamba, S.~M. Mullins, S.~H.~T. Murray, K.~P.
  Mutshena, A.~A. Pasternak, I.~Ragnarsson, D.~G. Roux, J.~F.
  Sharpey-Schafer, O.~Shirinda, P.~A. Vymers,
\newblock Phys. Lett. B 719 (2013) 83.

\bibitem{Masiteng2016DSAM}
P.~L. Masiteng, A.~A. Pasternak, E.~A. Lawrie, O.~Shirinda, J.~J.
  Lawrie, R.~A. Bark, S.~P. Bvumbi, N.~Y. Kheswa, R.~Lindsay, E.~O.
  Lieder, R.~M. Lieder, T.~E. Madiba, S.~M. Mullins, S.~H.~T.
  Murray, J.~Ndayishimye, S.~S. Ntshangase, P.~Papka, J.~F.
  Sharpey-Schafer,
\newblock Eur. Phys. J. A 52 (2016) 28.

\bibitem{Lawrie2008Possible}
E.~A. Lawrie, P.~A. Vymers, J.~J. Lawrie, Ch. Vieu, R.~A. Bark,
  R.~Lindsay, G.~K. Mabala, S.~M. Maliage, P.~L. Masiteng, S.~M.
  Mullins, S.~H.~T. Murray, I.~Ragnarsson, T.~M. Ramashidzha,
  C.~Sch\"uck, J.~F. Sharpey-Schafer, O.~Shirinda.
\newblock Phys. Rev. C 78 (2008) 021305(R).

\end{thebibliography}
\end{document}